\documentclass{llncs}
\usepackage{xcolor}
\usepackage{colortbl}
\usepackage{tikz,proof}
\usepackage{color}
\usepackage{gnuplot-lua-tikz}

\usepackage{xpatch}
\usepackage{tabu}

\makeatletter
\xpatchcmd{%
\@maketitle}{%
\ifx\@empty\authors \else \@setauthors \fi
}{%
  \ifx\@empty\authors \else \@setauthors \fi

  \ifx\@empty\addresses \else\@setaddresses\fi
}{\typeout{Patch successful}}{\typeout{Patch failed}}
\makeatother

\usepackage{amsmath}
\usepackage{amssymb}
\usepackage{mathtools}
\usepackage{hhline}

\usepackage[bookmarks=false]{hyperref}

\newcommand{\var}{\mathit{var}}
\newcommand{\true}{1}
\newcommand{\false}{0}

\newcommand{\emptys}{\bot}
\newcommand{\apply}[1]{{{\hspace{.1em}|\hspace{.075em}}\lower .25ex\hbox{$#1$}}}

\title{Trimming Graphs Using Clausal Proof Optimization}
\author{Marijn J.H. Heule}

\institute{Department of Computer Science, The University of Texas at Austin}

\def\L#1{\raise .2ex\hbox{\tiny\tt #1}&}

\def\K#1{\textbf{#1}}
\def\N{\\[.1ex]}                                      
\def\I{\hspace{1em}}
 
\begin{document}
\maketitle
\begin{abstract}
We present a method to gradually compute a smaller and smaller unsatisfiable core of a propositional formula by minimizing
proofs of unsatisfiability. The goal is to compute a minimal unsatisfiable core that is relatively small compared to
other minimal unsatisfiable cores of the same formula. We try to achieve this goal by postponing deletion of arbitrary clauses from the formula as long as
possible---in contrast to existing minimal unsatisfiable core algorithms. We applied this method to reduce the smallest
known unit-distance graph with chromatic number $5$ from $553$ vertices and $2\,720$ edges to $529$ vertices and $2\,670$ edges.

\end{abstract}

\section{Introduction}

Today's satisfiability (SAT) solvers can not only determine whether a propositional formula can be satisfied, but they
can also produce a certificate in case no satisfying assignments exists. These certificates, known as proofs of unsatisfiability,
can be used for multiple purposes ranging from checking the correctness of the unsatisfiability
claim~\cite{Heule:2014:delete,Goldberg,Gelder:2008,DBLP:conf/tacas/Cruz-FilipeMS17,ITP2017} to computing interpolants~\cite{DRUPing}. 
In this paper, we focus on another application of proofs of unsatisfiability: computing an unsatisfiable core of the
formula~\cite{Heule:2013:trim,Wetzler2014,ChildCheck,MUS2014}. We observed that the size of proofs tends to correlate to the
size the corresponding unsatisfiable cores: the smaller the proof, the smaller the unsatisfiable core.
We present a method to exploit this relation by computing a smaller and smaller proof of unsatisfiability to compute
a small unsatisfiable core. This method was developed to improve the upper bound of the smallest unit-distance graph with 
chromatic number 5, which is currently a Polymath project. Details about the problem and this project are described below.
The presented method was developed as existing techniques performed poorly on this application. Yet it could  
help with other applications that use unsatisfiable cores too---which we plan to study in the near future. 

The {chromatic number of the plane}, a problem first proposed by 
 Nelson in 1950~\cite{coloring},
asks how many colors are needed to color all points of the plane such that no two points at distance 1 from each other have the same color.
Early results showed that at least four and at most seven colors are required.
By the de Bruijn--Erd\H{o}s theorem, the chromatic number of the plane is the largest possible chromatic number
of a finite unit-distance graph~\cite{deBruijn}.
The Moser Spindle, a unit-distance graph with 7 vertices and 11 edges, shows the lower bound~\cite{Moser},
while the upper bound is shown by a 7-coloring of the entire plane by Isbell~\cite{coloring}.

In a breakthrough for this problem in April 2018, Aubrey de Grey improved the lower bound by providing
a unit-distance graph with $1\,581$ vertices with chromatic number 5~\cite{DeGrey}.
This discovery by de Grey started a Polymath project to find smaller graphs.
The current record is a graph with $553$ vertices and $2\,720$ edges~\cite{Geo}. We present a new  
technique to construct a large unit-distance graph with chromatic number 5, which we reduce with the proposed
method to a graph with ``only'' $529$ vertices and $2\,670$ edges.
This graph is much more symmetric compared to earlier small unit-distance graphs with chromatic number $5$.
The total costs to compute this graph were roughly $100\,000$ CPU hours.

\section{Preliminaries}

\paragraph{\bf Propositional Formulas.}
We will minimize graphs on the propositional level. 
We consider formulas in \emph{conjunctive normal form} (CNF), 
which are defined as follows. 
A \emph{literal} is either a variable $x$ (a \emph{positive literal}) 
or the negation $\overline x$ of a variable~$x$ (a \emph{negative literal}). 
The \emph{complement} $\overline l$ of a literal $l$ is defined as 
$\overline l = \overline x$ if $l = x$ and $\overline l = x$ if $l = \overline x$.
For a literal $l$, $\var(l)$ denotes the variable of $l$.
A \emph{clause} is a disjunction of literals and a \emph{formula} is a conjunction of clauses.

An \emph{assignment} is a function from a set of variables to the truth values 
\true{}~(\emph{true}) and \false{} (\emph{false}).
A literal $l$ is \emph{satisfied} by an assignment $\alpha$ if 
$l$ is positive and \mbox{$\alpha(\var(l)) = \true$} or if it is negative and $\alpha(\var(l)) = \false$.
A literal is \emph{falsified} by an assignment if its complement is satisfied by the assignment.
A clause is satisfied by an assignment $\alpha$ if it contains a literal that is satisfied by~$\alpha$.
A formula is satisfied by an assignment $\alpha$ if all its clauses are satisfied by $\alpha$.
A formula is \emph{satisfiable} if there exists an assignment that satisfies it and \emph{unsatisfiable} otherwise.

For a formula~$F$ and assignment $\alpha$, we denote by $F\apply{\alpha}$ a reduced copy of $F$ without
clauses satisfied by $\alpha$ and literals falsified by $\alpha$. 
A \emph{unit clause} is a clause with only one literal. 
The result of applying the \emph{unit clause rule} to a formula $F$ is the formula $F\apply{l}$ where $(l)$ is a unit clause in $F$.
The iterated application of this rule to a formula, until no unit clauses 
are left, is called \emph{unit propagation}. If unit propagation yields the empty clause $\emptys$, we
say that it derived a \emph{conflict}.

\paragraph{\bf Clausal Proofs.}

%
A clause $C$ is \emph{redundant} with respect to a formula $F$ if $F$ and $F \land C$ are satisfiability equivalent.
%
For instance, the clause $C = (x \lor y)$ is redundant with respect to the formula $F = (\overline x \lor \overline y)$ since $F$
and $F \land C$ are satisfiability equivalent (although they are not logically equivalent).
This redundancy notion allows us to add redundant clauses to a formula while preserving
satisfiability. 

Given a formula $F = \{C_1, \dots, C_m\}$, a \emph{clausal derivation} of a clause $C_n$ from $F$ is a sequence $C_{m+1}, \dots, C_n$ of clauses.
Such a sequence gives rise to \mbox{formulas} $F_m, F_{m+1}, \dots, F_n$, where $F_i = \{C_1, \dots, C_i\}$. We call $F_i$ the \emph{accumulated formula} corresponding to the \mbox{$i$-th} proof step.
A clausal derivation is \emph{correct} if every clause $C_i$ ($i > m$) is redundant with respect to the formula $F_{i-1}$ and if this redundancy can be checked in polynomial time with respect to the size of the proof.
A clausal derivation is a \emph{proof} of a formula $F$ if it derives the unsatisfiable empty clause. 
Clearly, since every clause-addition step preserves satisfiability, and since the empty clause
is always false, a proof of $F$ certifies the unsatisfiability of~$F$.

Checking the correctness of a clause $C_i$ in a derivation consists of computing a justification why $C_i$ is redundant with respect by formula $F_{i-1}$. 
The most commonly used method for this purpose is \emph{reverse unit propagation} (RUP): Let $\alpha$ be the assignment that falsifies all literals in $C_i$. 
Clause $C_i$ has the RUP property if and only if unit propagation on $F_{i-1}\apply\alpha$ results in a conflict.
In this case the justification of $C_i$ consists of all clauses that were required to derive the conflict. Clausal proofs that can be validated using 
this method are called RUP proofs. Most SAT-solving techniques can be compactly expressed as RUP.

\begin{example}
Consider the formula below consisting of 3 variables and 7 clauses:
$$
F := (\overline y \lor z) \land (x \lor z) \land (\overline x \lor y) \land (\overline x \lor \overline y) \land (x \lor \overline y) \land (y \lor \overline z) \land (x \lor \overline z)
$$

A clausal proof of $F$ is $\overline y, z, \emptys$. A justification of this proof is shown in Fig.~\ref{fig:justification}.
This justification shows that $\overline y$ and  $z$ do not depend on each other. As a consequence, swapping them results in another correct proof.
Notice that clause $(x \lor \overline z)$ is not used in this justification and it is thus not part of the core of $F$. 

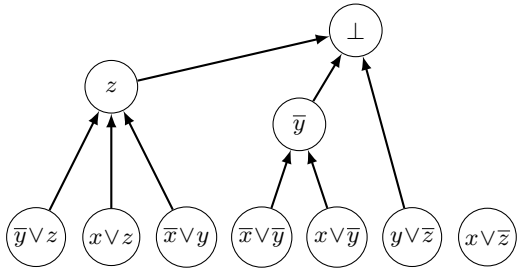
\begin{figure}[h]
\centering
 \tikzstyle{every circle node}=[circle,draw,inner sep=1.5pt,minimum size=0.7cm]
\tikzset{>=latex}

\begin{tikzpicture}
 \draw (0,0)        node[circle] (a)  {$\overline y  \!\lor\! z$};
 \draw (1,0)        node[circle] (b)  {$x \!\lor\! z$} ;
 \draw (2,0)        node[circle] (c)  {$\overline x \!\lor\! y$} ;
 \draw (3,0)        node[circle] (d)  {$\overline x \!\lor\! \overline y$} ;
 \draw (4,0)        node[circle] (e)  {$x \!\lor\! \overline y$} ;
 \draw (5,0)        node[circle] (f)  {$y \!\lor\! \overline z$} ;
 \draw (6,0)        node[circle] (g)  {$x \!\lor\! \overline z$} ;

 \draw (1,2)        node[circle] (aa) {$z$} ;
 \draw (3.5,1.5)    node[circle] (bb) {$\overline y$} ;
 \draw (4.25,2.75)  node[circle] (dd) {$\emptys$} ;

 \draw [thick,<-] (aa)--(a);
 \draw [thick,<-] (aa)--(b);
 \draw [thick,<-] (aa)--(c);

 \draw [thick,<-] (bb)--(d);
 \draw [thick,<-] (bb)--(e);


 \draw [thick,<-] (dd)--(aa);
 \draw [thick,<-] (dd)--(bb);
 \draw [thick,<-] (dd)--(f);

\end{tikzpicture}
\caption{A justification of the proof of the example formula. Each clause in the proof depends on its incoming arcs.
The clauses without incoming arcs represent the formula.}
\label{fig:justification}
\end{figure}

\end{example}

In practice, clausal proofs also contain deletion information. The presence of deletion information significantly reduces the
cost to compute a justification. Clausal proofs, which can be validated using the RUP method and include deletion information,
are known as DRUP proofs. We mostly ignore the deletion information aspect of clausal proofs
to simplify the presentation. All techniques discussed in this paper work with deletion information as well.

\paragraph{\bf Chromatic Number of the Plane.}

The Chromatic Number of the Plane (CNP)~\cite{coloring} asks how many colors are required in a coloring of the plane to
ensure that there exists no monochromatic pair of points with distance 1. A {\em unit-distance graph} is a graph formed from
a set of points in the plane by connecting two points by an edge whenever the distance between the two points is exactly one.
A lower bound for CNP of $k$ colors can be obtained by showing that a unit-distance graph has chromatic number $k$.

We will use three operations to construct larger and larger graphs:
the Minkowski sum~\cite{Hadwiger1950}, rotation, and merging.
Given two sets of points $A$ and $B$, the Minkowski sum of $A$ and $B$, denoted by $A \oplus B$, equals
\mbox{$\{a+b \mid a \in A, b \in B\}$}. Consider the sets of points $A = \{(0,0), (1,0)\}$ and \mbox{$B = \{(0,0), (1/2,\sqrt{3}/2)\}$}, then 
$A \oplus B = \{(0,0), (1,0), (1/2,\sqrt{3}/2), (3/2,\sqrt{3}/2)\}$.

\begin{figure}[t]
\vspace{-10pt}
\includegraphics[width=\textwidth]{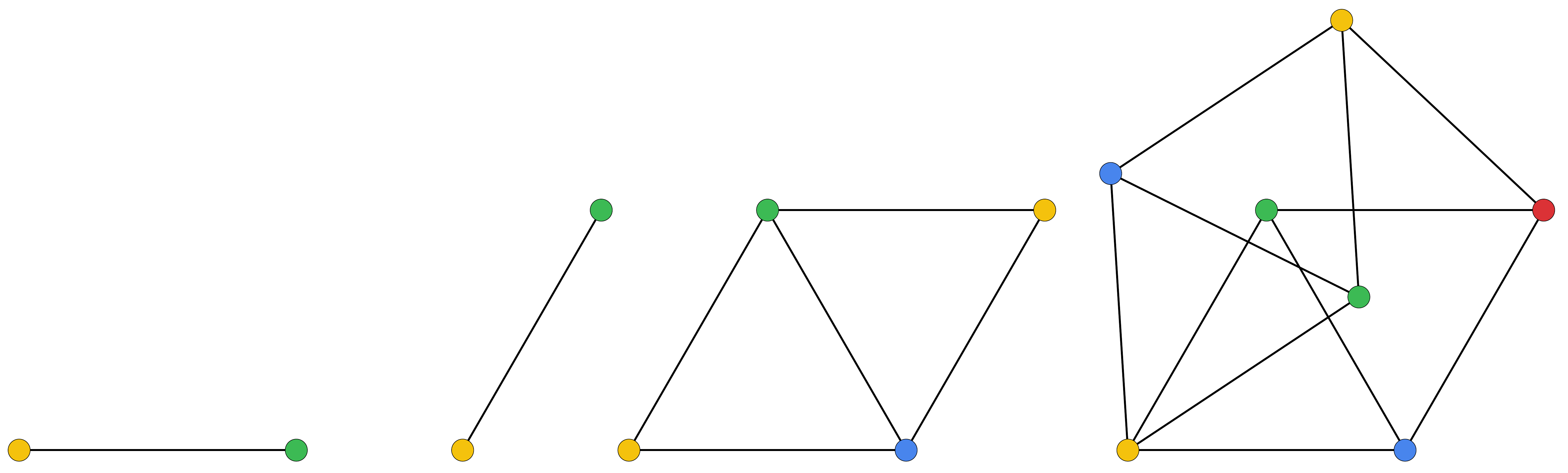}
\vspace{-10pt}
\caption{From left to right: illustrations of unit-distance graphs $A$, $B$, $A \oplus B$, and the Moser Spindle.
The graphs shown have chromatic number 2, 2, 3, and 4, respectively. The illustrations show valid colorings with the 
fewest number of colors.}
\vspace{-10pt}

\label{fig:intro}
\end{figure}

Given a positive integer $i$, we denote by $\theta_i$ the rotation around point $(0,0)$ with angle $\arccos(\frac{2i-1}{2i})$
and by $\theta_i^k$ the application of $\theta_i$ $k$ times.
Let $p$ be a point with distance $\sqrt{i}$ from $(0,0)$, then the
points $p$ and $\theta_i(p)$ are exactly distance 1 apart and thus would be connected with an 
edge in a unit-distance graph. Consider again the set of points
$A \oplus B$ above. 
The points $A \oplus B \cup \theta_3(A \oplus B)$ form the Moser Spindle~\cite{Moser} with chromatic number 4. 
Figure~\ref{fig:intro} shows visualizations of these sets with connected vertices colored differently.

\section{Overview of the Approach}

The smallest known unit-distance graph with chromatic number 5 has $553$ vertices and $2\,720$ edges~\cite{Geo}. This graph was found
using the following method. Start with a large unit-distance graph $G$ with chromatic number 5. Now reduce the size of that graph
by solving the formula that encodes whether the graph can be colored with 4 colors. That formula is unsatisfiable. From the proof
of unsatisfiability, an unsatisfiable core can be extracted that represents a subgraph with chromatic number 5. This step is 
repeated again and again as long as the graph is reduced. In the last step, vertices are randomly eliminated to make the
graph vertex-critical: removing any additional vertex introduces a 4-coloring. 

In this paper we present two improvements. The most important one is a new method, presented in Section~\ref{sec:min},
to produce short proofs of unsatisfiability. 
We observed that for the formulas studied in this paper that the shorter the proof, the smaller the unsatisfiable core and thus the smaller the subgraph. 
The second improvement is the construction of
a new large unit-distance graph $G$ that we use as a starting point to find a smaller unit-distance graph with chromatic number 5. The construction of this graph is 
explained in Section~\ref{sec:graph}. These two methods allowed us to find a unit-distance graph with $529$ vertices and $2\,670$ edges.
This graph is much more symmetric compared to the graph with $553$ vertices.

\section{Clausal Proof Optimization}
\label{sec:min}


Most SAT solvers can emit a clausal proof of unsatisfiability. There exist several checkers for such proofs, including formally-verified ones~\cite{Cruz-Filipe2017,Lammich2017}. 
We extended the
checker {\sf DRAT-trim}~\cite{Heule:2013:trim} that allows optimizing the clausal proof as well as extracting an unsatisfiable core. 
One can obtain multiple unsatisfiable cores from a single clausal proof---in contrast to a resolution proof~\cite{Zhang:2003}. 
The existing method works via backward checking~\cite{Goldberg}: Given a proof of unsatisfiability, the last clause (the empty clause) of the proof is marked.
Now the proof is validated in reverse order. For each marked clause it is determined which clauses (occurring earlier in the proof or in the 
formula) are required for the validation. Those clauses will be marked (if they were not marked already). The order in which unit propagation is applied 
influences which clauses become marked. Unmarked clauses are not validated. After the proof is verified, 
the marked clauses in the formula form an unsatisfiable core and the marked clauses in the proof form an optimized proof.
We present two new extensions that further reduce the size of the formula. 

\subsection{Justification Order Shuffling}

A clausal proof typically has many different justifications and a justification can typically be converted into many different
clausal proofs, i.e., clauses appear in a different order in the sequence. Here we exploit this property by 1) computing a
justification for a given clausal proof, 2) removing the clauses that are redundant based on that justification, and 3) shuffle
the remaining clauses in the proof based on that justification. These steps are repeated multiple times.

Figure~\ref{fig:opt} shows the pseudo code of that algorithm.
The procedure {\sf RemoveRedundancy} removes from a given clausal proof $P$ and justification $J$ all the 
clauses in $P$ that do not occur in any of the justifications of $J$. 
Given a justification $J$, the procedure {\sf ShuffleProof} produces a random permutation of the clauses in $J$ such that
each clause $C$ appears 1) later in the proof than all the clauses in the justification of $C$ and 2) before all clauses that list
$C$ in their justification. Additionally, {\sf ShuffleProof} randomly shuffles the literals of each clause in $J$. 

\begin{figure}[b]
\vspace{-10pt}
\centering
\begin{tabular}{r@{\quad}l}
\L{} {\sf OptimizeProof} (clausal proof $P$, formula $F$)  \N
\L{1} \I \K{do}\N
\L{2} \I\I $J$ := {\sf ComputeJustification} ($P$, $F$)   \N
\L{3} \I\I $J$ := {\sf RemoveRedundancy} ($J$)   \N
\L{4} \I\I $P$ := {\sf ShuffleProof} ($J$)   \N
\L{5} \I \K{while} (progress)\N
\L{6} \I\K{return} $P$ \N
\end{tabular}
\caption{Optimizing a proof by iterative computing a new justification.} 
\label{fig:opt}
\end{figure}

Clause deletion is not mentioned in the algorithm, but can also be helpful to optimize proofs. 
A clause $C$ can be deleted in a proof as soon as none of the clauses occurring later in the proof
uses $C$ in their justification. On the other hand, one could delete $C$ at a later point in the proof
(or not at all) to allow clauses later in the proof to incorporate $C$ in their justification in the
next iteration. We randomly postpone deleting clauses in proofs in a certain window. The 
window is slightly increased in each next iteration.  

\subsection{Iterative Trimming the Formula}
\label{sec:trim}

Given a unsatisfiable formula that encodes the existence of a $k$-coloring of a graph, an unsatisfiable
core of that formula represents a subgraph that cannot be colored with $k$ colors. To find a small subgraph we would
like a minimal unsatisfiable core and ideally the smallest minimal unsatisfiable core. Although there
has been some research in to the latter~\cite{SMUS1,SMUS2,SMUS3}, it is already hard to compute
a minimal unsatisfiable core. Existing algorithms for computing a minimal
unsatisfiable core~\cite{MUS2004,MUS2011} focus more on easy problems. For harder problems it is required to trim the formulas
using a preprocessing step~\cite{MUS2014}.




In preliminary experiments we observed that existing algorithms got stuck.
It turned out that if a ``wrong'' vertex is removed from the graph, then proving that the remaining 
graph still has chromatic number 5 is very expensive.
A proof that the initial graph has chromatic number 5 consists of roughly 10,000 clauses. 
After removing a clause that represents a ``wrong'' vertex, the proof consists of millions of clauses.
We concluded that existing tools are not effective for this application, because they
remove clauses arbitrary. This will eventually result in removing a clause 
representating a ``wrong'' vertex. 
Although the checking costs are a serious problem, there is a more 
problematic issue: as soon as it requires millions of clauses 
to prove that the graph has chromatic number five, then many vertices 
are involved in the proof and the minimal unsatisfiable core will be relatively large. As a consequence, 
this also holds for the graph represented by this core.
We address this issue by taking away the elimination of arbitrary clauses.
Instead, we only remove clauses via trimming and proof optimization.

\begin{figure}[b]
\centering
\vspace{-10pt}
\begin{tabular}{r@{\quad}l}
\L{} {\sf TrimFormulaPlain} (formula $F$)  \N
\L{1} \I $F_{\mathrm{core}} := F$ \N
\L{2} \I \K{do}\N
\L{3} \I\I $P$ := {\sf ComputeProof} ($F_{\mathrm{core}}$)   \N
\L{4} \I\I $P$ := {\sf OptimizeProof} ($P$, $F_{\mathrm{core}}$)   \N
\L{5} \I\I $F_{\mathrm{core}}$ := {\sf ComputeCore} ($P$, $F_{\mathrm{core}}$)   \N
\L{6} \I \K{while} (progress)\N
\L{7} \I\K{return} $F_{\mathrm{core}}$ \N
\end{tabular}
~~~~~
\begin{tabular}{r@{\quad}l}
\L{} {\sf TrimFormulaInteract} (formula $F$)  \N
\L{1} \I $F_{\mathrm{core}} := F$ \N
\L{2} \I \K{do}\N
\L{3} \I\I $P$ := {\sf ComputeProof} ($F_{\mathrm{core}}$)   \N
\L{4} \I\I $P$ := {\sf OptimizeProof} ($P$, $F_{\mathrm{core}}$)   \N
\L{5} \I\I $P$ := {\sf OptimizeProof} ($P$, $F$)   \N
\L{6} \I\I $F_{\mathrm{core}}$ := {\sf ComputeCore} ($P$, $F$)   \N
\L{7} \I \K{while} (progress)\N
\L{8} \I\K{return} $F_{\mathrm{core}}$ \N
\end{tabular}
\caption{Pseudo code of two algorithms to trim the size of a formula using proof optimization: {\sf TrimFormulaPlain} 
and {\sf TrimFormulaInteract}. The latter algorithm interacts with the original formula to further optimize the proof.} 
\label{fig:trim}
\end{figure}

Figure~\ref{fig:trim} shows the pseudo codes of two algorithms to trim a formula: one algorithm, called
{\sf TrimFormulaPlain}, that simply adds proof optimization to the trimming loop and another one, 
called {\sf TrimFormulaInteract}, that additionally interacts with the original formula to further optimize 
the proof. We focus on the latter algorithm, which is one of the main contributions of this paper.

Algorithm {\sf TrimFormulaInteract} takes advantage of the following property of (D)RUP proofs: If (D)RUP proof $P$ is
a correct proof of formula $F$, then $P$ is a correct proof of any formula $F'$ such that
$F' \supseteq F$. Observe that additional clauses cannot break 
the RUP check: if unit propagation on $F$ results in a conflict, then unit propagation on $F'$ results
in a conflict. 

In each step of the main loop of {\sf TrimFormulaInteract}, we first compute a proof of unsatisfiability of the
trimmed formula $F_{\mathrm{core}}$. The size of this proof is crucial for the quality
of the trimming. One could therefore solve $F_{\mathrm{core}}$ multiple times by shuffling the clauses
and select the smallest proof of these runs. Afterwards, this proof is optimized using $F_{\mathrm{core}}$
via the algorithm shown in Fig.~\ref{fig:opt}. Next, we use the property discussed above and 
further optimize the proof using $F$ and the same optimization algorithm. The algorithm has now more options
to minimize the proof as $F \supseteq F_{\mathrm{core}}$. Moreover, the algorithm allows for a novel way to 
compute a smaller core: In an earlier step a clause may have been removed that allows for a small
proof of unsatisfiability and/or small unsatisfiable core. Since each step considers again all clauses of $F$, that clause
may be pulled back into $F_{\mathrm{core}}$.

The size of $F_{\mathrm{core}}$ does not necessarily decrease with each iteration and may actually increase
if a low quality proof is computed in line 3. We repeat the algorithm as long as there is progress.
In this case, we measured progress by the reduction of the size of $F_{\mathrm{core}}$.

The result of these trimming algorithms is rarely a minimal unsatisfiable core of the formula. We applied the 
classical destructive method~\cite{deconstruct} to reduce $F_{\mathrm{core}}$ to a minimal unsatisfiable core. We observed
(some details are presented in Section~\ref{sec:large})
that the size of the minimal unsatisfiable core can vary significantly based on the selection of the clauses to 
remove. As a consequence we ran this method multiple (thousands of) times on the cluster to obtain a relatively
small minimal unsatisfiable core of $F_{\mathrm{core}}$.





\section{Observed patterns of points in $\mathbb{Q}[\sqrt{3}, \sqrt{11}] \times \mathbb{Q}[\sqrt{3}, \sqrt{11}]$}
\label{sec:graph}

The smallest known unit-distance graph with chromatic number 5, called $G_{553}$, has 553 vertices~\cite{Geo}. Its key component is 
a set of $420$ points embedded in $\mathbb{Q}[\sqrt{3}, \sqrt{11}] \times \mathbb{Q}[\sqrt{3}, \sqrt{11}]$ that have a limited 
number ($19$) of the colorings of the points at distance 2 from the origin (central vertex) when coloring the set with
4 colors. Our strategy to compute a small unit-distance graph with chromatic number 5 is finding a
small set of vertices with the same property. We explored many large graphs with points in 
$\mathbb{Q}[\sqrt{3}, \sqrt{11}] \times \mathbb{Q}[\sqrt{3}, \sqrt{11}]$ and computed the size of proofs of unsatisfiability of the
formula that determines the existence of a 4-coloring while blocking the limited 
number of the colorings of the points at distance $2$. This section describes how we obtained the large 
graph with the smallest proof of unsatisfiability that we encountered. 

\begin{figure}[h]
\centering
\includegraphics[width=0.35\textwidth]{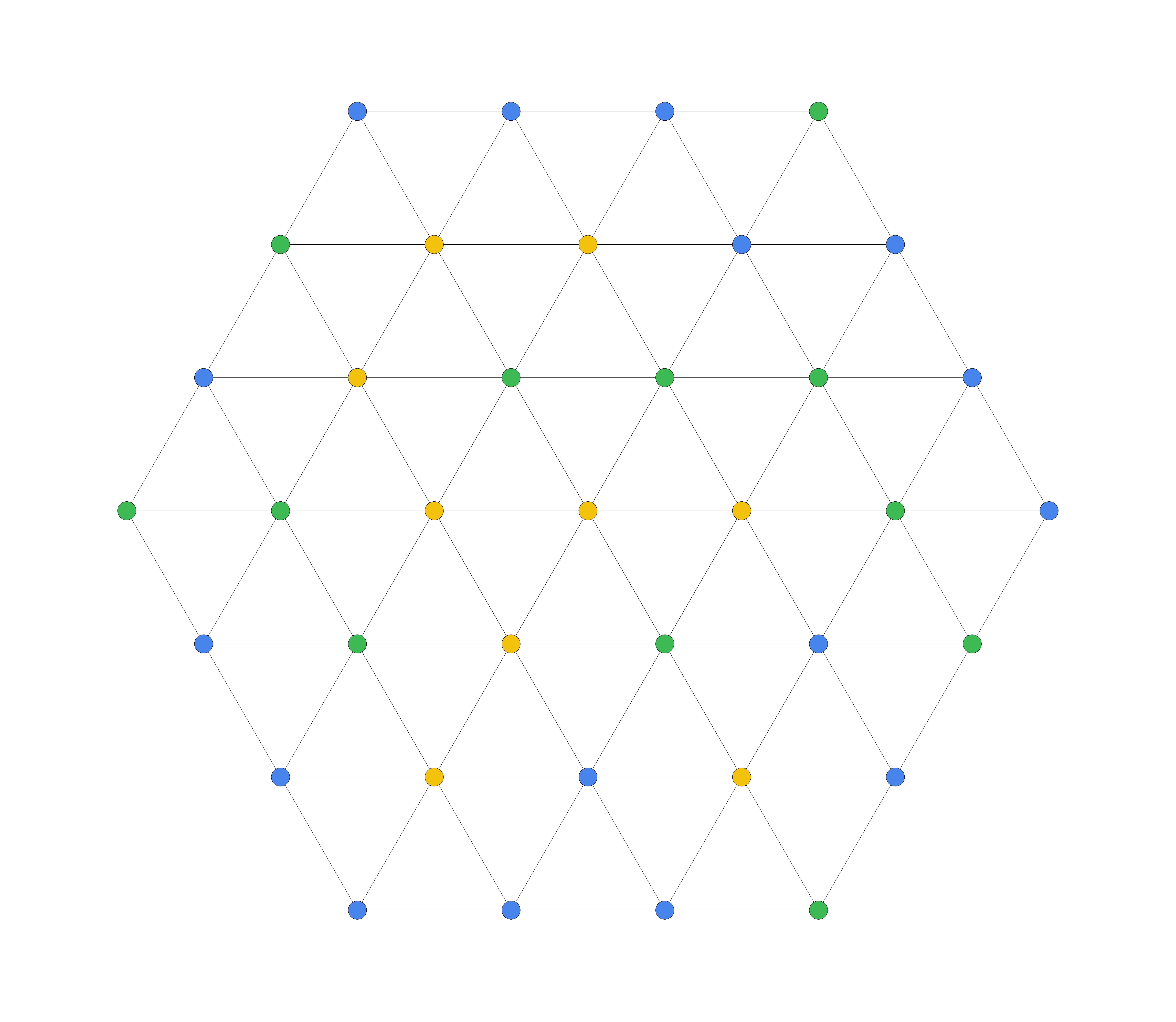}
\includegraphics[width=0.62\textwidth]{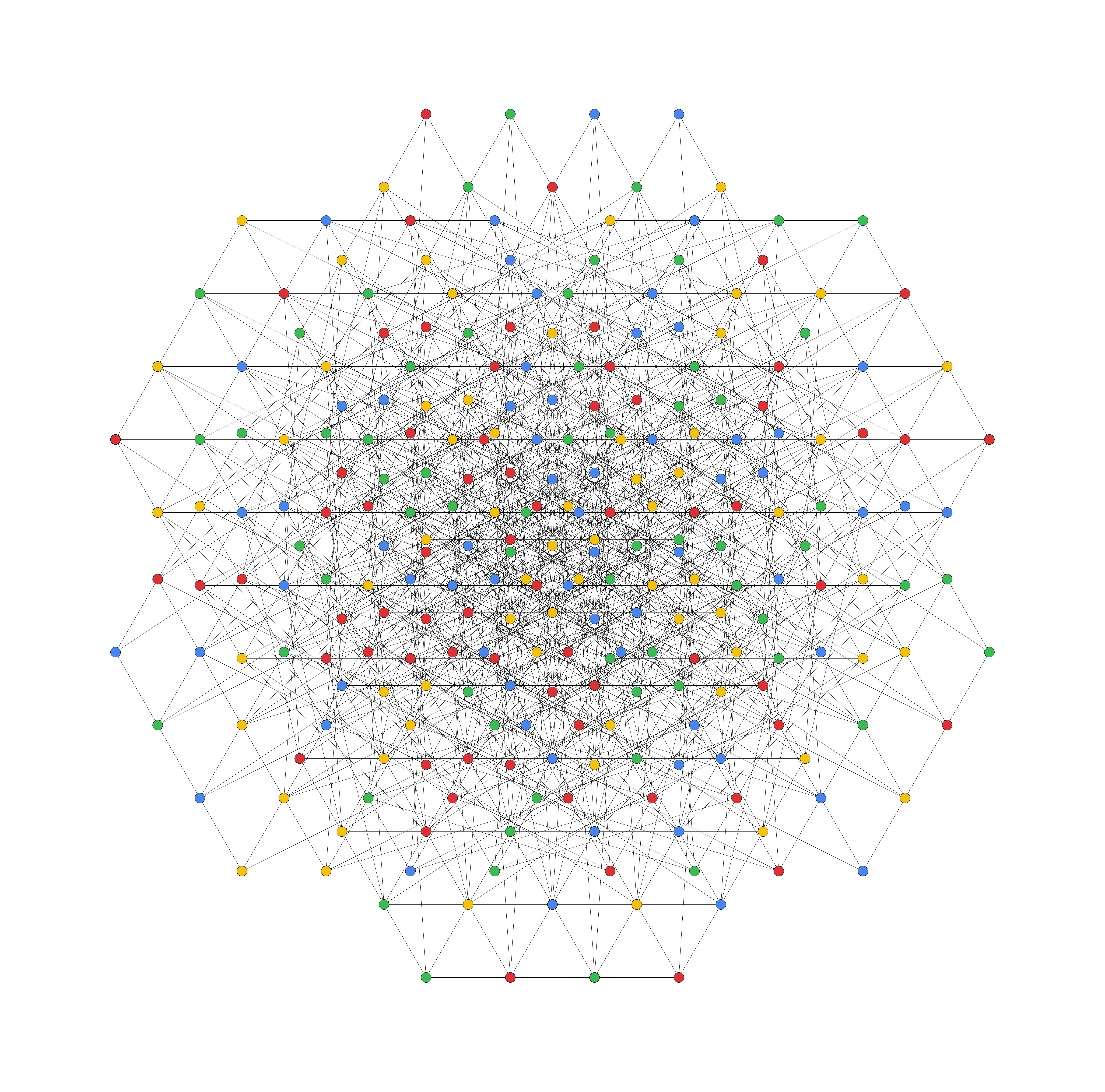}
\caption{A 3-coloring of the graph $H_{\frac{1}{3}} \oplus H_{\frac{1}{3}}  \oplus H_{\frac{1}{3}}$ (left) and a 4-coloring of
the graph $H_{\frac{1}{3}} \oplus H_{\frac{1}{3}}  \oplus H_{\frac{1}{3}} \oplus H'$\raisebox{2pt}{\!\!$_{{\frac{\sqrt{3}+\sqrt{11}}{6}}}$} (right).}
\label{fig:Honce}
\end{figure}

We denote by $H_R$ the graph consisting of i) a regular hexagon with maximal radius $R$ and ii) its center. 
The points of $H_R$ in the plane are $(0,0)$, $(R,0)$, $(R/2, R\sqrt{3}/2)$, $(-R/2, R\sqrt{3}/2)$, $(-R,0)$,
$(-R/2, -R\sqrt{3}/2)$ and $(R/2, -R\sqrt{3}/2)$. Furthermore, we denote by $H'_R$ a copy of $H_R$ rotated by 90 degrees.

We observed some interesting patterns when combining the graphs $H_{\frac{1}{3}}$ and  $H'$\raisebox{2pt}{\!\!$_{{\frac{\sqrt{3}+\sqrt{11}}{6}}}$}.
Figure~\ref{fig:Honce} (left) shows the graph $H_{\frac{1}{3}} \oplus H_{\frac{1}{3}}  \oplus H_{\frac{1}{3}}$, which is a triangular
grid with diameter $1$. This graph has $37$ vertices and 48 edges and can be colored with $3$ colors. However, the Minkowski sum of this triangular grid and
$H'$\raisebox{2pt}{\!\!$_{{\frac{\sqrt{3}+\sqrt{11}}{6}}}$}, 
shown in Fig.~\ref{fig:Honce} (right), is not 3-colorable. Notice that there are many edges between the seven triangular grids. Actually, 
the graph has $259$ vertices and $1\,056$ edges and most of these edges ($720$) are between triangular grids. There exist many 4-colorings
of this graph and most of them have no observable pattern. 

\begin{figure}[h]
\centering
\includegraphics[width=.98\textwidth]{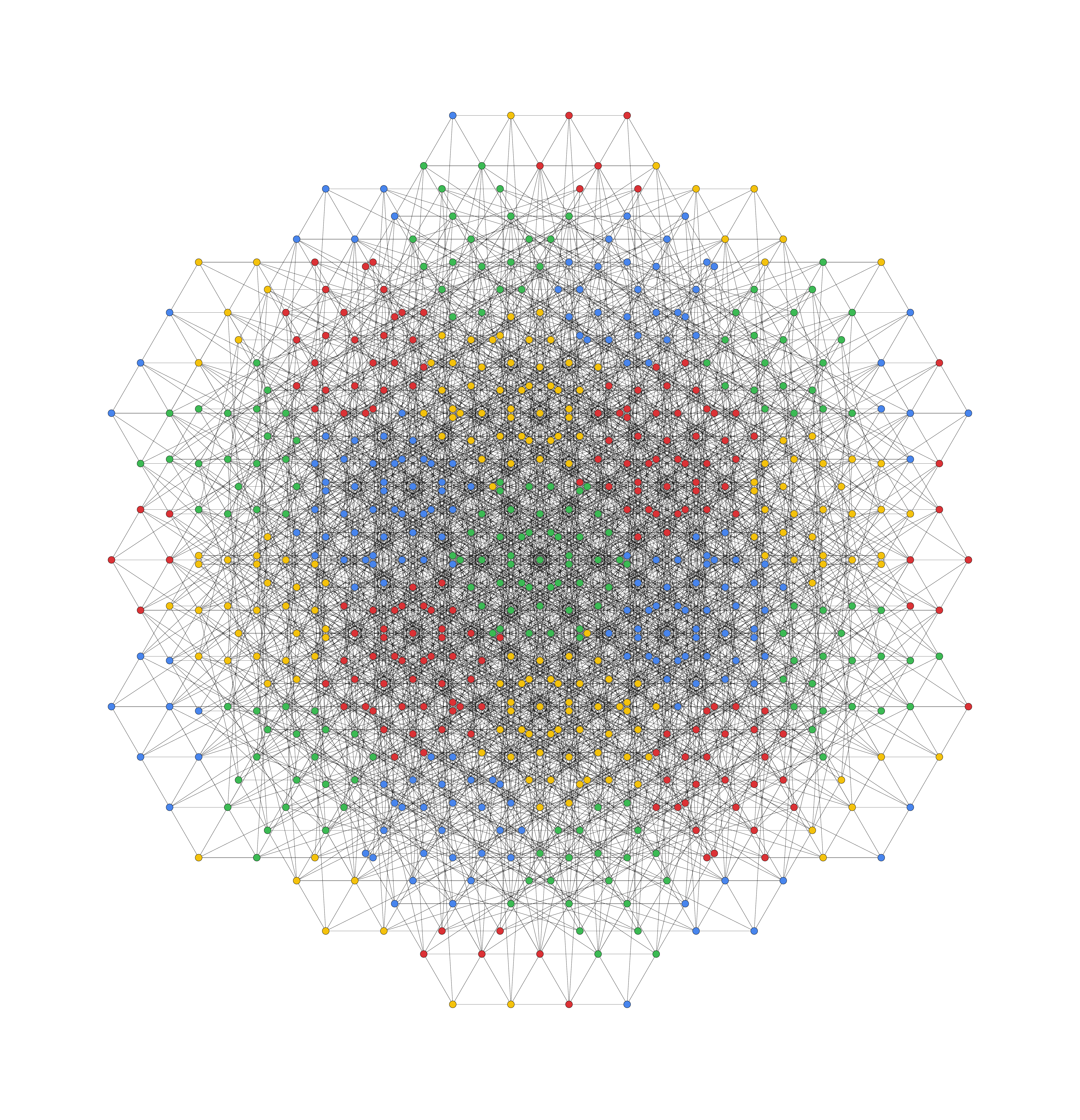}
\caption{A 4-coloring of the graph $H_{\frac{1}{3}} \oplus H_{\frac{1}{3}}  \oplus H_{\frac{1}{3}} \oplus  H'$\raisebox{2pt}{\!\!$_{{\frac{\sqrt{3}+\sqrt{11}}{6}}}$}$ \oplus  H'$\raisebox{2pt}{\!\!$_{{\frac{\sqrt{3}+\sqrt{11}}{6}}}$}.}
\label{fig:Htwice}
\end{figure}

Patterns start to emerge when applying the Minkowski sum again. Figure~\ref{fig:Htwice} shows a 4-coloring of the resulting 
graph $H_{\frac{1}{3}} \oplus H_{\frac{1}{3}}  \oplus H_{\frac{1}{3}} \oplus  H'$\raisebox{2pt}{\!\!$_{{\frac{\sqrt{3}+\sqrt{11}}{6}}}$}$ \oplus H'$\raisebox{2pt}{\!\!$_{{\frac{\sqrt{3}+\sqrt{11}}{6}}}$}. Observe the clustering of vertices with the same color in circles of roughly a diameter
of 1 in size. This pattern can be observed in many of the found 4-colorings of this graph, although there also exist some 4-colorings without this pattern. It appears that assigning the same color to nearby vertices is the easiest way to color this graph (using a SAT solver). 

\begin{figure}[h!]
\centering
\includegraphics[width=.9\textwidth]{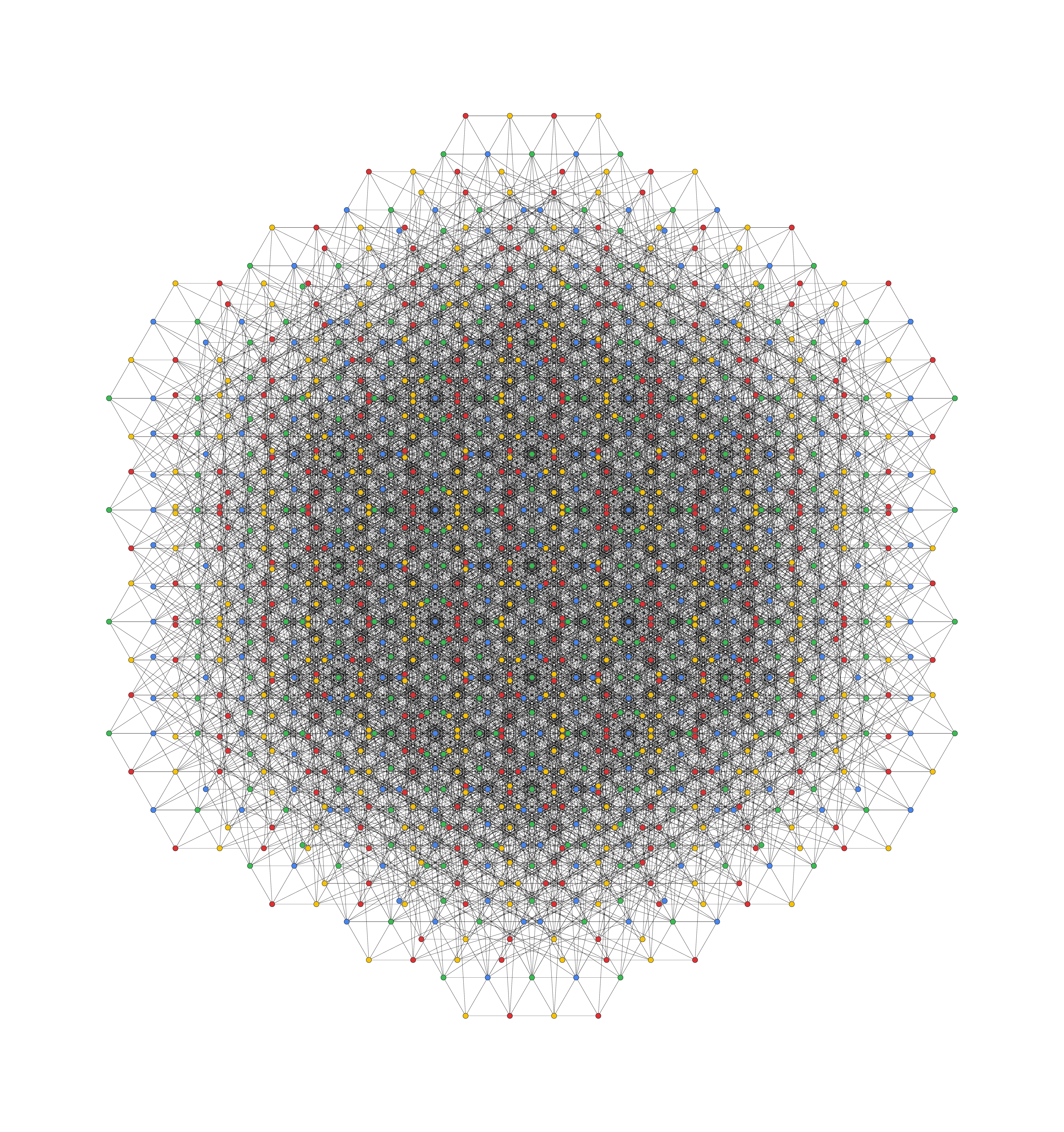}
\caption{A 4-coloring of the graph $H_{\frac{1}{3}} \oplus H_{\frac{1}{3}}  \oplus H_{\frac{1}{3}} \oplus H'$\raisebox{2pt}{\!\!$_{{\frac{\sqrt{3}+\sqrt{11}}{6}}}$}$ \oplus H'$\raisebox{2pt}{\!\!$_{{\frac{\sqrt{3}+\sqrt{11}}{6}}}$}$\oplus H'$\raisebox{2pt}{\!\!$_{{\frac{\sqrt{3}+\sqrt{11}}{6}}}$}.}
\label{fig:Htriple}
\end{figure}

Applying the Minkowski sum another time breaks the prior pattern completely, as no more 4-colorings exist with clusters of vertices
having the same color. However, new patterns emerge, as can be seen in Fig~\ref{fig:Htriple}. For example, notice the reflection in 
the central vertical axis of the blue and green vertices.

\begin{figure}[h]
\centering
\includegraphics[width=.98\textwidth]{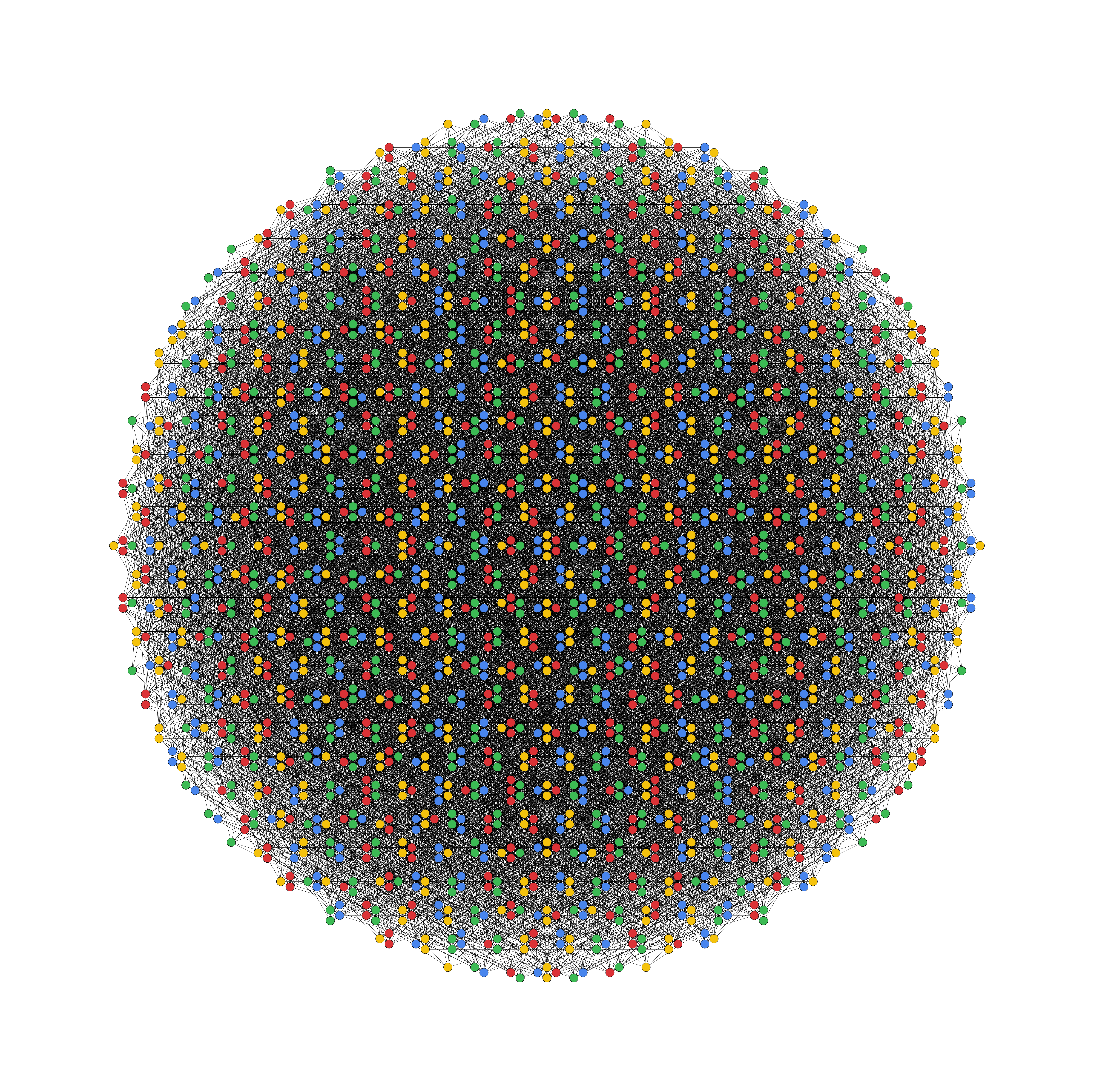}
\caption{A 4-coloring of graph $G_{2167}$.}
\label{fig:2167}
\end{figure}

Based on these observations, we experimented with ways to combine $H_{\frac{1}{3}}$ and $H'$\raisebox{2pt}{\!\!$_{{\frac{\sqrt{3}+\sqrt{11}}{6}}}$}. An effective 
combination turned out to be unit-distance graph $G_{2167}$. This graph is constructed as follows.
Let $C_{13}$ denote the union of  $H_{\frac{1}{3}}$ and $H'$\raisebox{2pt}{\!\!$_{{\frac{\sqrt{3}+\sqrt{11}}{6}}}$}. Now $G_{2167}$
equals $C_{13} \oplus C_{13} \oplus C_{13} \oplus C_{13} \oplus C_{13} \oplus C_{13} \oplus C_{13} \oplus C_{13}$ without 
the points that have a distance larger than 2 from the central vertex. This graph has $2\,167$ vertices and $16\,512$ edges and is shown in Fig.~\ref{fig:2167}.
Notice that the average vertex degree is larger than 15. This is quite high for a graph with chromatic number 4. 

Observe the vertical monochromatic lines in Fig.~\ref{fig:2167}: Points with the same horizontal coordinate have the same color. 
This pattern appears in many $4$-colorings (modulo a rotation of $60$ degrees). There are solutions with vertical lines with two colors, 
but none of the $4$-colorings have more colors on a single vertical line (again, modulo
a rotation of $60$ degrees). The only reason why such solutions can exist is that the construction of $G_{2167}$ does not 
generate points with distance 1 that have the same horizontal coordinate. There appears no obvious way to add such 
points in a way that the resulting graph has chromatic number 5. Another pattern that can be observed in Fig.~\ref{fig:2167}
is that points with the same vertical coordinate that are 2/3 apart from each other also have the same color. Also any two
points that are 1/3 apart have a different color. For example, forcing that any vertex at distance 1/3 from the origin has the
same color as the central vertex eliminates all 4-colorings. Hence 1/3 is a so-called virtual-edge in 4-colorings of unit-distance 
graphs. 

\section{Small Unit-Distance Graph with Chromatic Number 5}

In this section we present our SAT-based approach to improve the smallest known unit-distance graph with chromatic number 5.
We first explain how we encode the problem and afterwards apply the new trimming algorithm presented in Section~\ref{sec:trim}.


\subsection{Encoding}

We can compute the chromatic number of a graph $G$ as follows. Construct two formulas,  
one asking whether $G$ can be colored with $k-1$ colors, and one whether $G$ can be colored with $k$ colors. Now, 
$G$ has chromatic number $k$ if and only if the former is unsatisfiable while the latter is satisfiable. 

The construction of these two formulas can be achieved using the following encoding~\cite{Geo}: Given a graph $G = (V,E)$
and a parameter $k$, the encoding uses $k|V|$ boolean variables $x_{v,c}$ with $v \in V$ and $c \in \{1,\dots,k\}$.
These variables have the following meaning: $x_{v,c}$ is true if and only if vertex $v$ has color $c$. Now we can construct
a propositional formula $F_k$ that is satisfiable if and only if $G$ can be colored with $k$ colors:
\[
F_k : = \bigwedge_{v \in V} (x_{v,1} \lor \dots \lor x_{v,k}) \land \bigwedge_{\{v,w\} \in E} \bigwedge_{c \in \{1,\dots,k\}} (\overline x_{v,c} \lor \overline x_{w,c})
\]

The first type of clauses, called vertex clauses, ensures that each vertex has at least one color, while the second type of clauses, called edge clauses,
forces that two connected vertices are colored differently. Additionally, we could include clauses to require that each vertex has at most one color. However, these
clauses are redundant and would be eliminated by blocked clause elimination~\cite{BCE}, a SAT preprocessing technique. We experimented
using formulas with and without blocked clauses. Although the results were quite similar, we had the impression that without blocked clauses 
is slightly better.

We added symmetry-breaking predicates~\cite{Crawford} during all experiments to speed up solving and proof minimization. 
The color symmetries were broken by fixing the vertex at ($0$, $0$) to the first color, the vertex at ($1$, $0$) to the second color, and the vertex
at ($1/2$, $\sqrt{3}/2$) to the third color. These three points are at distance 1 from each other and occurred in all our graphs. The speedup is roughly a factor of
24 ($= 4 \cdot 3 \cdot 2)$, when proving the absence of a 4-coloring.

\begin{figure}[t]
\hspace{-12pt}
\begin{minipage}{.48\textwidth}
\begin{tikzpicture}[gnuplot,scale=0.52]
\path (0.000,0.000) rectangle (12.500,8.750);
\gpcolor{color=gp lt color border}
\gpsetlinetype{gp lt border}
\gpsetdashtype{gp dt solid}
\gpsetlinewidth{1.00}
\draw[gp path] (1.564,0.616)--(1.744,0.616);
\draw[gp path] (11.947,0.616)--(11.767,0.616);
\node[gp node right] at (1.780,0.616) {$10^3$};
\draw[gp path] (1.564,1.785)--(1.654,1.785);
\draw[gp path] (11.947,1.785)--(11.857,1.785);
\draw[gp path] (1.564,2.468)--(1.654,2.468);
\draw[gp path] (11.947,2.468)--(11.857,2.468);
\draw[gp path] (1.564,2.953)--(1.654,2.953);
\draw[gp path] (11.947,2.953)--(11.857,2.953);
\draw[gp path] (1.564,3.330)--(1.654,3.330);
\draw[gp path] (11.947,3.330)--(11.857,3.330);
\draw[gp path] (1.564,3.637)--(1.654,3.637);
\draw[gp path] (11.947,3.637)--(11.857,3.637);
\draw[gp path] (1.564,3.897)--(1.654,3.897);
\draw[gp path] (11.947,3.897)--(11.857,3.897);
\draw[gp path] (1.564,4.122)--(1.654,4.122);
\draw[gp path] (11.947,4.122)--(11.857,4.122);
\draw[gp path] (1.564,4.321)--(1.654,4.321);
\draw[gp path] (11.947,4.321)--(11.857,4.321);
\draw[gp path] (1.564,4.499)--(1.744,4.499);
\draw[gp path] (11.947,4.499)--(11.767,4.499);
\node[gp node right] at (1.780,4.499) {$10^4$};
\draw[gp path] (1.564,5.667)--(1.654,5.667);
\draw[gp path] (11.947,5.667)--(11.857,5.667);
\draw[gp path] (1.564,6.351)--(1.654,6.351);
\draw[gp path] (11.947,6.351)--(11.857,6.351);
\draw[gp path] (1.564,6.836)--(1.654,6.836);
\draw[gp path] (11.947,6.836)--(11.857,6.836);
\draw[gp path] (1.564,7.212)--(1.654,7.212);
\draw[gp path] (11.947,7.212)--(11.857,7.212);
\draw[gp path] (1.564,7.520)--(1.654,7.520);
\draw[gp path] (11.947,7.520)--(11.857,7.520);
\draw[gp path] (1.564,7.780)--(1.654,7.780);
\draw[gp path] (11.947,7.780)--(11.857,7.780);
\draw[gp path] (1.564,8.005)--(1.654,8.005);
\draw[gp path] (11.947,8.005)--(11.857,8.005);
\draw[gp path] (1.564,8.203)--(1.654,8.203);
\draw[gp path] (11.947,8.203)--(11.857,8.203);
\draw[gp path] (1.564,8.381)--(1.744,8.381);
\draw[gp path] (11.947,8.381)--(11.767,8.381);
\node[gp node right] at (1.780,8.381) {$10^5$};
\draw[gp path] (1.564,0.616)--(1.564,0.796);
\draw[gp path] (1.564,8.381)--(1.564,8.201);
\node[gp node center] at (1.564,0.308) {$0$};
\draw[gp path] (4.160,0.616)--(4.160,0.796);
\draw[gp path] (4.160,8.381)--(4.160,8.201);
\node[gp node center] at (4.160,0.308) {$5$};
\draw[gp path] (6.756,0.616)--(6.756,0.796);
\draw[gp path] (6.756,8.381)--(6.756,8.201);
\node[gp node center] at (6.756,0.308) {$10$};
\draw[gp path] (9.351,0.616)--(9.351,0.796);
\draw[gp path] (9.351,8.381)--(9.351,8.201);
\node[gp node center] at (9.351,0.308) {$15$};
\draw[gp path] (11.947,0.616)--(11.947,0.796);
\draw[gp path] (11.947,8.381)--(11.947,8.201);
\node[gp node center] at (11.947,0.308) {$20$};
\draw[gp path] (1.564,8.381)--(1.564,0.616)--(11.947,0.616)--(11.947,8.381)--cycle;
\node[gp node right] at (10.479,8.047) {size of unsatisfiable core};
\gpcolor{rgb color={0.580,0.000,0.827}}
\draw[gp path] (10.663,8.047)--(11.579,8.047);
\draw[gp path] (1.564,7.737)--(2.083,5.025)--(2.602,4.960)--(3.121,4.934)--(3.641,4.923)%
  --(4.160,4.918)--(4.679,4.910)--(5.198,4.902)--(5.717,4.918)--(6.236,4.891)--(6.756,4.873)%
  --(7.275,4.892)--(7.794,4.876)--(8.313,4.877)--(8.832,4.881)--(9.351,4.898)--(9.870,4.845)%
  --(10.390,4.834)--(10.909,4.839)--(11.428,4.830)--(11.947,4.832);
\gpsetpointsize{4.00}
\gppoint{gp mark 1}{(1.564,7.737)}
\gppoint{gp mark 1}{(2.083,5.025)}
\gppoint{gp mark 1}{(2.602,4.960)}
\gppoint{gp mark 1}{(3.121,4.934)}
\gppoint{gp mark 1}{(3.641,4.923)}
\gppoint{gp mark 1}{(4.160,4.918)}
\gppoint{gp mark 1}{(4.679,4.910)}
\gppoint{gp mark 1}{(5.198,4.902)}
\gppoint{gp mark 1}{(5.717,4.918)}
\gppoint{gp mark 1}{(6.236,4.891)}
\gppoint{gp mark 1}{(6.756,4.873)}
\gppoint{gp mark 1}{(7.275,4.892)}
\gppoint{gp mark 1}{(7.794,4.876)}
\gppoint{gp mark 1}{(8.313,4.877)}
\gppoint{gp mark 1}{(8.832,4.881)}
\gppoint{gp mark 1}{(9.351,4.898)}
\gppoint{gp mark 1}{(9.870,4.845)}
\gppoint{gp mark 1}{(10.390,4.834)}
\gppoint{gp mark 1}{(10.909,4.839)}
\gppoint{gp mark 1}{(11.428,4.830)}
\gppoint{gp mark 1}{(11.947,4.832)}
\gppoint{gp mark 1}{(11.121,8.047)}
\gpcolor{color=gp lt color border}
\node[gp node right] at (10.479,7.439) {size of proof of unsatisfiability};
\gpcolor{rgb color={0.000,0.620,0.451}}
\draw[gp path] (10.663,7.439)--(11.579,7.439);
\draw[gp path] (1.564,1.616)--(2.083,1.390)--(2.602,1.382)--(3.121,1.381)--(3.641,1.378)%
  --(4.160,1.371)--(4.679,1.367)--(5.198,1.359)--(5.717,1.356)--(6.236,1.353)--(6.756,1.352)%
  --(7.275,1.351)--(7.794,1.345)--(8.313,1.345)--(8.832,1.344)--(9.351,1.343)--(9.870,1.340)%
  --(10.390,1.337)--(10.909,1.335)--(11.428,1.335)--(11.947,1.335);
\gppoint{gp mark 2}{(1.564,1.616)}
\gppoint{gp mark 2}{(2.083,1.390)}
\gppoint{gp mark 2}{(2.602,1.382)}
\gppoint{gp mark 2}{(3.121,1.381)}
\gppoint{gp mark 2}{(3.641,1.378)}
\gppoint{gp mark 2}{(4.160,1.371)}
\gppoint{gp mark 2}{(4.679,1.367)}
\gppoint{gp mark 2}{(5.198,1.359)}
\gppoint{gp mark 2}{(5.717,1.356)}
\gppoint{gp mark 2}{(6.236,1.353)}
\gppoint{gp mark 2}{(6.756,1.352)}
\gppoint{gp mark 2}{(7.275,1.351)}
\gppoint{gp mark 2}{(7.794,1.345)}
\gppoint{gp mark 2}{(8.313,1.345)}
\gppoint{gp mark 2}{(8.832,1.344)}
\gppoint{gp mark 2}{(9.351,1.343)}
\gppoint{gp mark 2}{(9.870,1.340)}
\gppoint{gp mark 2}{(10.390,1.337)}
\gppoint{gp mark 2}{(10.909,1.335)}
\gppoint{gp mark 2}{(11.428,1.335)}
\gppoint{gp mark 2}{(11.947,1.335)}
\gppoint{gp mark 2}{(11.121,7.439)}
\gpcolor{color=gp lt color border}
\draw[gp path] (1.564,8.381)--(1.564,0.616)--(11.947,0.616)--(11.947,8.381)--cycle;
\gpdefrectangularnode{gp plot 1}{\pgfpoint{1.564cm}{0.616cm}}{\pgfpoint{11.947cm}{8.381cm}}
\end{tikzpicture}
\end{minipage}
~\,
\begin{minipage}{.48\textwidth}
\begin{tikzpicture}[gnuplot,scale=0.52]
\path (0.000,0.000) rectangle (12.500,8.750);
\gpcolor{color=gp lt color border}
\gpsetlinetype{gp lt border}
\gpsetdashtype{gp dt solid}
\gpsetlinewidth{1.00}
\draw[gp path] (1.564,0.616)--(1.744,0.616);
\draw[gp path] (11.947,0.616)--(11.767,0.616);
\node[gp node right] at (1.780,0.616) {$10^3$};
\draw[gp path] (1.564,1.785)--(1.654,1.785);
\draw[gp path] (11.947,1.785)--(11.857,1.785);
\draw[gp path] (1.564,2.468)--(1.654,2.468);
\draw[gp path] (11.947,2.468)--(11.857,2.468);
\draw[gp path] (1.564,2.953)--(1.654,2.953);
\draw[gp path] (11.947,2.953)--(11.857,2.953);
\draw[gp path] (1.564,3.330)--(1.654,3.330);
\draw[gp path] (11.947,3.330)--(11.857,3.330);
\draw[gp path] (1.564,3.637)--(1.654,3.637);
\draw[gp path] (11.947,3.637)--(11.857,3.637);
\draw[gp path] (1.564,3.897)--(1.654,3.897);
\draw[gp path] (11.947,3.897)--(11.857,3.897);
\draw[gp path] (1.564,4.122)--(1.654,4.122);
\draw[gp path] (11.947,4.122)--(11.857,4.122);
\draw[gp path] (1.564,4.321)--(1.654,4.321);
\draw[gp path] (11.947,4.321)--(11.857,4.321);
\draw[gp path] (1.564,4.499)--(1.744,4.499);
\draw[gp path] (11.947,4.499)--(11.767,4.499);
\node[gp node right] at (1.780,4.499) {$10^4$};
\draw[gp path] (1.564,5.667)--(1.654,5.667);
\draw[gp path] (11.947,5.667)--(11.857,5.667);
\draw[gp path] (1.564,6.351)--(1.654,6.351);
\draw[gp path] (11.947,6.351)--(11.857,6.351);
\draw[gp path] (1.564,6.836)--(1.654,6.836);
\draw[gp path] (11.947,6.836)--(11.857,6.836);
\draw[gp path] (1.564,7.212)--(1.654,7.212);
\draw[gp path] (11.947,7.212)--(11.857,7.212);
\draw[gp path] (1.564,7.520)--(1.654,7.520);
\draw[gp path] (11.947,7.520)--(11.857,7.520);
\draw[gp path] (1.564,7.780)--(1.654,7.780);
\draw[gp path] (11.947,7.780)--(11.857,7.780);
\draw[gp path] (1.564,8.005)--(1.654,8.005);
\draw[gp path] (11.947,8.005)--(11.857,8.005);
\draw[gp path] (1.564,8.203)--(1.654,8.203);
\draw[gp path] (11.947,8.203)--(11.857,8.203);
\draw[gp path] (1.564,8.381)--(1.744,8.381);
\draw[gp path] (11.947,8.381)--(11.767,8.381);
\node[gp node right] at (1.780,8.381) {$10^5$};
\draw[gp path] (1.564,0.616)--(1.564,0.796);
\draw[gp path] (1.564,8.381)--(1.564,8.201);
\node[gp node center] at (1.564,0.308) {$0$};
\draw[gp path] (4.160,0.616)--(4.160,0.796);
\draw[gp path] (4.160,8.381)--(4.160,8.201);
\node[gp node center] at (4.160,0.308) {$5$};
\draw[gp path] (6.756,0.616)--(6.756,0.796);
\draw[gp path] (6.756,8.381)--(6.756,8.201);
\node[gp node center] at (6.756,0.308) {$10$};
\draw[gp path] (9.351,0.616)--(9.351,0.796);
\draw[gp path] (9.351,8.381)--(9.351,8.201);
\node[gp node center] at (9.351,0.308) {$15$};
\draw[gp path] (11.947,0.616)--(11.947,0.796);
\draw[gp path] (11.947,8.381)--(11.947,8.201);
\node[gp node center] at (11.947,0.308) {$20$};
\draw[gp path] (1.564,8.381)--(1.564,0.616)--(11.947,0.616)--(11.947,8.381)--cycle;
\node[gp node right] at (10.479,8.047) {size of unsatisfiable core};
\gpcolor{rgb color={0.580,0.000,0.827}}
\draw[gp path] (10.663,8.047)--(11.579,8.047);
\draw[gp path] (1.564,7.737)--(2.083,6.333)--(2.602,6.312)--(3.121,6.307)--(3.641,6.293)%
  --(4.160,6.294)--(4.679,6.288)--(5.198,6.286)--(5.717,6.271)--(6.236,6.273)--(6.756,6.257)%
  --(7.275,6.260)--(7.794,6.251)--(8.313,6.271)--(8.832,6.264)--(9.351,6.254)--(9.870,6.258)%
  --(10.390,6.262)--(10.909,6.274)--(11.428,6.257)--(11.947,6.262);
\gpsetpointsize{4.00}
\gppoint{gp mark 1}{(1.564,7.737)}
\gppoint{gp mark 1}{(2.083,6.333)}
\gppoint{gp mark 1}{(2.602,6.312)}
\gppoint{gp mark 1}{(3.121,6.307)}
\gppoint{gp mark 1}{(3.641,6.293)}
\gppoint{gp mark 1}{(4.160,6.294)}
\gppoint{gp mark 1}{(4.679,6.288)}
\gppoint{gp mark 1}{(5.198,6.286)}
\gppoint{gp mark 1}{(5.717,6.271)}
\gppoint{gp mark 1}{(6.236,6.273)}
\gppoint{gp mark 1}{(6.756,6.257)}
\gppoint{gp mark 1}{(7.275,6.260)}
\gppoint{gp mark 1}{(7.794,6.251)}
\gppoint{gp mark 1}{(8.313,6.271)}
\gppoint{gp mark 1}{(8.832,6.264)}
\gppoint{gp mark 1}{(9.351,6.254)}
\gppoint{gp mark 1}{(9.870,6.258)}
\gppoint{gp mark 1}{(10.390,6.262)}
\gppoint{gp mark 1}{(10.909,6.274)}
\gppoint{gp mark 1}{(11.428,6.257)}
\gppoint{gp mark 1}{(11.947,6.262)}
\gppoint{gp mark 1}{(11.121,8.047)}
\gpcolor{color=gp lt color border}
\node[gp node right] at (10.479,7.439) {size of proof of unsatisfiability};
\gpcolor{rgb color={0.000,0.620,0.451}}
\draw[gp path] (10.663,7.439)--(11.579,7.439);
\draw[gp path] (1.564,7.207)--(2.083,7.000)--(2.602,6.990)--(3.121,6.985)--(3.641,6.981)%
  --(4.160,6.979)--(4.679,6.976)--(5.198,6.974)--(5.717,6.971)--(6.236,6.966)--(6.756,6.961)%
  --(7.275,6.957)--(7.794,6.945)--(8.313,6.942)--(8.832,6.938)--(9.351,6.935)--(9.870,6.933)%
  --(10.390,6.930)--(10.909,6.929)--(11.428,6.926)--(11.947,6.924);
\gppoint{gp mark 2}{(1.564,7.207)}
\gppoint{gp mark 2}{(2.083,7.000)}
\gppoint{gp mark 2}{(2.602,6.990)}
\gppoint{gp mark 2}{(3.121,6.985)}
\gppoint{gp mark 2}{(3.641,6.981)}
\gppoint{gp mark 2}{(4.160,6.979)}
\gppoint{gp mark 2}{(4.679,6.976)}
\gppoint{gp mark 2}{(5.198,6.974)}
\gppoint{gp mark 2}{(5.717,6.971)}
\gppoint{gp mark 2}{(6.236,6.966)}
\gppoint{gp mark 2}{(6.756,6.961)}
\gppoint{gp mark 2}{(7.275,6.957)}
\gppoint{gp mark 2}{(7.794,6.945)}
\gppoint{gp mark 2}{(8.313,6.942)}
\gppoint{gp mark 2}{(8.832,6.938)}
\gppoint{gp mark 2}{(9.351,6.935)}
\gppoint{gp mark 2}{(9.870,6.933)}
\gppoint{gp mark 2}{(10.390,6.930)}
\gppoint{gp mark 2}{(10.909,6.929)}
\gppoint{gp mark 2}{(11.428,6.926)}
\gppoint{gp mark 2}{(11.947,6.924)}
\gppoint{gp mark 2}{(11.121,7.439)}
\gpcolor{color=gp lt color border}
\draw[gp path] (1.564,8.381)--(1.564,0.616)--(11.947,0.616)--(11.947,8.381)--cycle;
\gpdefrectangularnode{gp plot 1}{\pgfpoint{1.564cm}{0.616cm}}{\pgfpoint{11.947cm}{8.381cm}}
\end{tikzpicture}
\end{minipage}
\vspace{-10pt}
\caption{The size (number of clauses) of the unsatisfiable core and the optimized proof of unsatisfiability (y-axis) of the first twenty steps (x-axis) of the {\sf OptimizeProof} algorithm,
when starting with $F^+_{4}$ and the smallest proof (left) or the largest proof (right).}
\label{fig:seeds}
\end{figure}
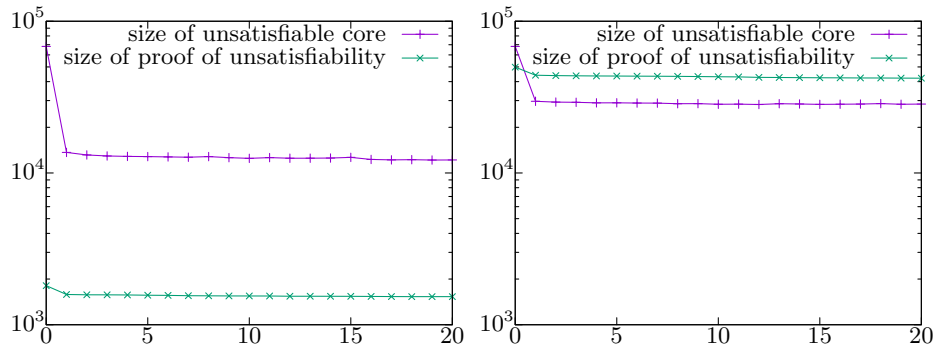

\subsection{Reducing the Large Part}
\label{sec:large}

The smallest known unit-distance graph with chromatic number $5$ has $553$ vertices and consists of two parts: a large part with $420$ vertices 
and a small part with $134$ vertices. The large part and small part have one vertex in common: the origin. Analysis of these parts~\cite{Geo}
showed that they have different purposes: the large part limits the number of valid 4-coloring of $12$ vertices at distance 2 from the origin to $19$. 
The small part prevents these $12$ vertices to having any of these $19$ 4-colorings. Some important details are missing from this analysis and
they will be discussed later. We focused our effort to search for a small unit-distance
graph with chromatic number 5 by looking for a more compact large part.

In the first step, we constructed the formula whether graph $G_{2167}$ has a 4-coloring. Apart from the symmetry-breaking predicates, 
we added $19$ clauses that block the above mentioned 4-colorings that remain in the large part. This formula, called $F^+_{4}$, is unsatisfiable and has
$8\,668$ variables and $68\,237$ clauses. In the next step we produce a proof of unsatisfiability of this formula. We used the SAT solver
{\sf glucose} 3.0~\cite{glucose} (without preprocessing techniques) for this purpose. 
This solver allows to randomly initialize the decision heuristics (VSIDS), which is a feature that can easily be added to most SAT solvers. This initialization
can have a significant impact on the size of the proof \emph{and} on the size of the core. For example, we solved the formula with 100
different seeds for the initialization. The smallest proof had $1\,809$ clause addition steps, while the largest proof had $49\,838$
clause addition steps. The default {\sf glucose} 3.0, i.e., without decision heuristics initialization, produced a proof with $2\,475$ clause 
addition steps.

Figure~\ref{fig:seeds} shows the effect of using the smallest and largest proof as input for the {\sf OptimizeProof} algorithm, which has been 
implemented in the {\sf DRAT-trim} proof checker (available at \url{https://github.com/marijnheule/drat-trim})~\cite{Heule:2013:trim}. In both cases
the size of the proof reduction is modest. However, a much smaller unsatisfiable core can be extracted from the optimized smallest proof
compared to the optimized largest proof. The smaller core also corresponds to a smaller subgraph ($963$ versus $1\,609$ vertices).

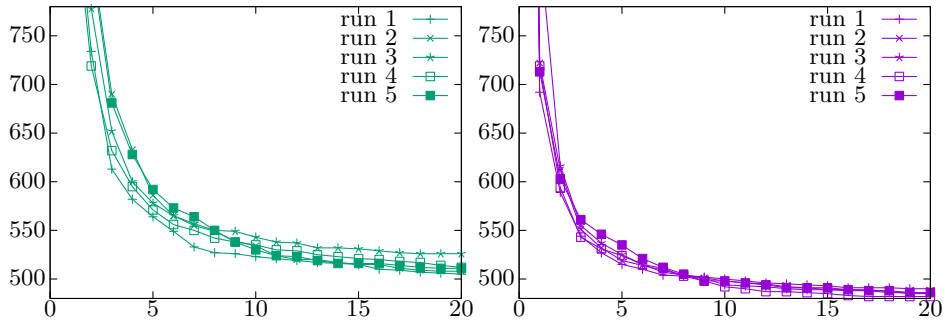
\begin{figure}[b]
\hspace{-9pt}
\begin{minipage}{.48\textwidth}
\begin{tikzpicture}[gnuplot,scale=0.5]
\path (0.000,0.000) rectangle (12.500,8.750);
\gpcolor{color=gp lt color border}
\gpsetlinetype{gp lt border}
\gpsetdashtype{gp dt solid}
\gpsetlinewidth{1.00}
\draw[gp path] (1.012,1.134)--(1.192,1.134);
\draw[gp path] (11.947,1.134)--(11.767,1.134);
\node[gp node right] at (1.128,1.134) {$500$};
\draw[gp path] (1.012,2.428)--(1.192,2.428);
\draw[gp path] (11.947,2.428)--(11.767,2.428);
\node[gp node right] at (1.128,2.428) {$550$};
\draw[gp path] (1.012,3.722)--(1.192,3.722);
\draw[gp path] (11.947,3.722)--(11.767,3.722);
\node[gp node right] at (1.128,3.722) {$600$};
\draw[gp path] (1.012,5.016)--(1.192,5.016);
\draw[gp path] (11.947,5.016)--(11.767,5.016);
\node[gp node right] at (1.128,5.016) {$650$};
\draw[gp path] (1.012,6.310)--(1.192,6.310);
\draw[gp path] (11.947,6.310)--(11.767,6.310);
\node[gp node right] at (1.128,6.310) {$700$};
\draw[gp path] (1.012,7.605)--(1.192,7.605);
\draw[gp path] (11.947,7.605)--(11.767,7.605);
\node[gp node right] at (1.128,7.605) {$750$};
\draw[gp path] (1.012,0.616)--(1.012,0.796);
\draw[gp path] (1.012,8.381)--(1.012,8.201);
\node[gp node center] at (1.012,0.308) {$0$};
\draw[gp path] (3.746,0.616)--(3.746,0.796);
\draw[gp path] (3.746,8.381)--(3.746,8.201);
\node[gp node center] at (3.746,0.308) {$5$};
\draw[gp path] (6.480,0.616)--(6.480,0.796);
\draw[gp path] (6.480,8.381)--(6.480,8.201);
\node[gp node center] at (6.480,0.308) {$10$};
\draw[gp path] (9.213,0.616)--(9.213,0.796);
\draw[gp path] (9.213,8.381)--(9.213,8.201);
\node[gp node center] at (9.213,0.308) {$15$};
\draw[gp path] (11.947,0.616)--(11.947,0.796);
\draw[gp path] (11.947,8.381)--(11.947,8.201);
\node[gp node center] at (11.947,0.308) {$20$};
\draw[gp path] (1.012,8.381)--(1.012,0.616)--(11.947,0.616)--(11.947,8.381)--cycle;
\node[gp node right] at (10.479,8.047) {run 1};
\gpcolor{rgb color={0.000,0.620,0.451}}
\draw[gp path] (10.663,8.047)--(11.579,8.047);
\draw[gp path] (1.992,8.381)--(2.106,7.190)--(2.652,4.058)--(3.199,3.256)--(3.746,2.790)%
  --(4.293,2.402)--(4.839,1.988)--(5.386,1.833)--(5.933,1.807)--(6.480,1.729)--(7.026,1.677)%
  --(7.573,1.625)--(8.120,1.574)--(8.667,1.548)--(9.213,1.522)--(9.760,1.393)--(10.307,1.367)%
  --(10.854,1.315)--(11.400,1.289)--(11.947,1.263);
\gpsetpointsize{4.00}
\gppoint{gp mark 1}{(2.106,7.190)}
\gppoint{gp mark 1}{(2.652,4.058)}
\gppoint{gp mark 1}{(3.199,3.256)}
\gppoint{gp mark 1}{(3.746,2.790)}
\gppoint{gp mark 1}{(4.293,2.402)}
\gppoint{gp mark 1}{(4.839,1.988)}
\gppoint{gp mark 1}{(5.386,1.833)}
\gppoint{gp mark 1}{(5.933,1.807)}
\gppoint{gp mark 1}{(6.480,1.729)}
\gppoint{gp mark 1}{(7.026,1.677)}
\gppoint{gp mark 1}{(7.573,1.625)}
\gppoint{gp mark 1}{(8.120,1.574)}
\gppoint{gp mark 1}{(8.667,1.548)}
\gppoint{gp mark 1}{(9.213,1.522)}
\gppoint{gp mark 1}{(9.760,1.393)}
\gppoint{gp mark 1}{(10.307,1.367)}
\gppoint{gp mark 1}{(10.854,1.315)}
\gppoint{gp mark 1}{(11.400,1.289)}
\gppoint{gp mark 1}{(11.947,1.263)}
\gppoint{gp mark 1}{(11.121,8.047)}
\gpcolor{color=gp lt color border}
\node[gp node right] at (10.479,7.539) {run 2};
\gpcolor{rgb color={0.000,0.620,0.451}}
\draw[gp path] (10.663,7.539)--(11.579,7.539);
\draw[gp path] (2.315,8.381)--(2.652,6.052)--(3.199,4.576)--(3.746,3.360)--(4.293,2.816)%
  --(4.839,2.557)--(5.386,2.402)--(5.933,2.143)--(6.480,2.014)--(7.026,1.755)--(7.573,1.651)%
  --(8.120,1.651)--(8.667,1.574)--(9.213,1.522)--(9.760,1.522)--(10.307,1.418)--(10.854,1.367)%
  --(11.400,1.341)--(11.947,1.341);
\gppoint{gp mark 2}{(2.652,6.052)}
\gppoint{gp mark 2}{(3.199,4.576)}
\gppoint{gp mark 2}{(3.746,3.360)}
\gppoint{gp mark 2}{(4.293,2.816)}
\gppoint{gp mark 2}{(4.839,2.557)}
\gppoint{gp mark 2}{(5.386,2.402)}
\gppoint{gp mark 2}{(5.933,2.143)}
\gppoint{gp mark 2}{(6.480,2.014)}
\gppoint{gp mark 2}{(7.026,1.755)}
\gppoint{gp mark 2}{(7.573,1.651)}
\gppoint{gp mark 2}{(8.120,1.651)}
\gppoint{gp mark 2}{(8.667,1.574)}
\gppoint{gp mark 2}{(9.213,1.522)}
\gppoint{gp mark 2}{(9.760,1.522)}
\gppoint{gp mark 2}{(10.307,1.418)}
\gppoint{gp mark 2}{(10.854,1.367)}
\gppoint{gp mark 2}{(11.400,1.341)}
\gppoint{gp mark 2}{(11.947,1.341)}
\gppoint{gp mark 2}{(11.121,7.539)}
\gpcolor{color=gp lt color border}
\node[gp node right] at (10.479,7.031) {run 3};
\gpcolor{rgb color={0.000,0.620,0.451}}
\draw[gp path] (10.663,7.031)--(11.579,7.031);
\draw[gp path] (2.100,8.381)--(2.106,8.329)--(2.652,5.068)--(3.199,3.722)--(3.746,3.153)%
  --(4.293,2.816)--(4.839,2.609)--(5.386,2.428)--(5.933,2.402)--(6.480,2.247)--(7.026,2.117)%
  --(7.573,2.091)--(8.120,1.962)--(8.667,1.962)--(9.213,1.936)--(9.760,1.884)--(10.307,1.833)%
  --(10.854,1.807)--(11.400,1.807)--(11.947,1.807);
\gppoint{gp mark 3}{(2.106,8.329)}
\gppoint{gp mark 3}{(2.652,5.068)}
\gppoint{gp mark 3}{(3.199,3.722)}
\gppoint{gp mark 3}{(3.746,3.153)}
\gppoint{gp mark 3}{(4.293,2.816)}
\gppoint{gp mark 3}{(4.839,2.609)}
\gppoint{gp mark 3}{(5.386,2.428)}
\gppoint{gp mark 3}{(5.933,2.402)}
\gppoint{gp mark 3}{(6.480,2.247)}
\gppoint{gp mark 3}{(7.026,2.117)}
\gppoint{gp mark 3}{(7.573,2.091)}
\gppoint{gp mark 3}{(8.120,1.962)}
\gppoint{gp mark 3}{(8.667,1.962)}
\gppoint{gp mark 3}{(9.213,1.936)}
\gppoint{gp mark 3}{(9.760,1.884)}
\gppoint{gp mark 3}{(10.307,1.833)}
\gppoint{gp mark 3}{(10.854,1.807)}
\gppoint{gp mark 3}{(11.400,1.807)}
\gppoint{gp mark 3}{(11.947,1.807)}
\gppoint{gp mark 3}{(11.121,7.031)}
\gpcolor{color=gp lt color border}
\node[gp node right] at (10.479,6.523) {run 4};
\gpcolor{rgb color={0.000,0.620,0.451}}
\draw[gp path] (10.663,6.523)--(11.579,6.523);
\draw[gp path] (1.942,8.381)--(2.106,6.802)--(2.652,4.550)--(3.199,3.593)--(3.746,2.971)%
  --(4.293,2.583)--(4.839,2.428)--(5.386,2.221)--(5.933,2.117)--(6.480,2.040)--(7.026,1.910)%
  --(7.573,1.884)--(8.120,1.781)--(8.667,1.729)--(9.213,1.677)--(9.760,1.651)--(10.307,1.600)%
  --(10.854,1.574)--(11.400,1.496)--(11.947,1.444);
\gppoint{gp mark 4}{(2.106,6.802)}
\gppoint{gp mark 4}{(2.652,4.550)}
\gppoint{gp mark 4}{(3.199,3.593)}
\gppoint{gp mark 4}{(3.746,2.971)}
\gppoint{gp mark 4}{(4.293,2.583)}
\gppoint{gp mark 4}{(4.839,2.428)}
\gppoint{gp mark 4}{(5.386,2.221)}
\gppoint{gp mark 4}{(5.933,2.117)}
\gppoint{gp mark 4}{(6.480,2.040)}
\gppoint{gp mark 4}{(7.026,1.910)}
\gppoint{gp mark 4}{(7.573,1.884)}
\gppoint{gp mark 4}{(8.120,1.781)}
\gppoint{gp mark 4}{(8.667,1.729)}
\gppoint{gp mark 4}{(9.213,1.677)}
\gppoint{gp mark 4}{(9.760,1.651)}
\gppoint{gp mark 4}{(10.307,1.600)}
\gppoint{gp mark 4}{(10.854,1.574)}
\gppoint{gp mark 4}{(11.400,1.496)}
\gppoint{gp mark 4}{(11.947,1.444)}
\gppoint{gp mark 4}{(11.121,6.523)}
\gpcolor{color=gp lt color border}
\node[gp node right] at (10.479,6.015) {run 5};
\gpcolor{rgb color={0.000,0.620,0.451}}
\draw[gp path] (10.663,6.015)--(11.579,6.015);
\draw[gp path] (2.284,8.381)--(2.652,5.819)--(3.199,4.447)--(3.746,3.515)--(4.293,3.023)%
  --(4.839,2.790)--(5.386,2.428)--(5.933,2.117)--(6.480,1.910)--(7.026,1.755)--(7.573,1.729)%
  --(8.120,1.625)--(8.667,1.548)--(9.213,1.548)--(9.760,1.548)--(10.307,1.496)--(10.854,1.444)%
  --(11.400,1.418)--(11.947,1.418);
\gppoint{gp mark 5}{(2.652,5.819)}
\gppoint{gp mark 5}{(3.199,4.447)}
\gppoint{gp mark 5}{(3.746,3.515)}
\gppoint{gp mark 5}{(4.293,3.023)}
\gppoint{gp mark 5}{(4.839,2.790)}
\gppoint{gp mark 5}{(5.386,2.428)}
\gppoint{gp mark 5}{(5.933,2.117)}
\gppoint{gp mark 5}{(6.480,1.910)}
\gppoint{gp mark 5}{(7.026,1.755)}
\gppoint{gp mark 5}{(7.573,1.729)}
\gppoint{gp mark 5}{(8.120,1.625)}
\gppoint{gp mark 5}{(8.667,1.548)}
\gppoint{gp mark 5}{(9.213,1.548)}
\gppoint{gp mark 5}{(9.760,1.548)}
\gppoint{gp mark 5}{(10.307,1.496)}
\gppoint{gp mark 5}{(10.854,1.444)}
\gppoint{gp mark 5}{(11.400,1.418)}
\gppoint{gp mark 5}{(11.947,1.418)}
\gppoint{gp mark 5}{(11.121,6.015)}
\gpcolor{color=gp lt color border}
\draw[gp path] (1.012,8.381)--(1.012,0.616)--(11.947,0.616)--(11.947,8.381)--cycle;
\gpdefrectangularnode{gp plot 1}{\pgfpoint{1.012cm}{0.616cm}}{\pgfpoint{11.947cm}{8.381cm}}
\end{tikzpicture}
\end{minipage}
~\,
\begin{minipage}{.48\textwidth}
\begin{tikzpicture}[gnuplot,scale=0.5]
\path (0.000,0.000) rectangle (12.500,8.750);
\gpcolor{color=gp lt color border}
\gpsetlinetype{gp lt border}
\gpsetdashtype{gp dt solid}
\gpsetlinewidth{1.00}
\draw[gp path] (1.012,1.134)--(1.192,1.134);
\draw[gp path] (11.947,1.134)--(11.767,1.134);
\node[gp node right] at (1.128,1.134) {$500$};
\draw[gp path] (1.012,2.428)--(1.192,2.428);
\draw[gp path] (11.947,2.428)--(11.767,2.428);
\node[gp node right] at (1.128,2.428) {$550$};
\draw[gp path] (1.012,3.722)--(1.192,3.722);
\draw[gp path] (11.947,3.722)--(11.767,3.722);
\node[gp node right] at (1.128,3.722) {$600$};
\draw[gp path] (1.012,5.016)--(1.192,5.016);
\draw[gp path] (11.947,5.016)--(11.767,5.016);
\node[gp node right] at (1.128,5.016) {$650$};
\draw[gp path] (1.012,6.310)--(1.192,6.310);
\draw[gp path] (11.947,6.310)--(11.767,6.310);
\node[gp node right] at (1.128,6.310) {$700$};
\draw[gp path] (1.012,7.605)--(1.192,7.605);
\draw[gp path] (11.947,7.605)--(11.767,7.605);
\node[gp node right] at (1.128,7.605) {$750$};
\draw[gp path] (1.012,0.616)--(1.012,0.796);
\draw[gp path] (1.012,8.381)--(1.012,8.201);
\node[gp node center] at (1.012,0.308) {$0$};
\draw[gp path] (3.746,0.616)--(3.746,0.796);
\draw[gp path] (3.746,8.381)--(3.746,8.201);
\node[gp node center] at (3.746,0.308) {$5$};
\draw[gp path] (6.480,0.616)--(6.480,0.796);
\draw[gp path] (6.480,8.381)--(6.480,8.201);
\node[gp node center] at (6.480,0.308) {$10$};
\draw[gp path] (9.213,0.616)--(9.213,0.796);
\draw[gp path] (9.213,8.381)--(9.213,8.201);
\node[gp node center] at (9.213,0.308) {$15$};
\draw[gp path] (11.947,0.616)--(11.947,0.796);
\draw[gp path] (11.947,8.381)--(11.947,8.201);
\node[gp node center] at (11.947,0.308) {$20$};
\draw[gp path] (1.012,8.381)--(1.012,0.616)--(11.947,0.616)--(11.947,8.381)--cycle;
\node[gp node right] at (10.479,8.047) {run 1};
\gpcolor{rgb color={0.580,0.000,0.827}}
\draw[gp path] (10.663,8.047)--(11.579,8.047);
\draw[gp path] (1.526,8.381)--(1.559,6.103)--(2.106,3.437)--(2.652,2.273)--(3.199,1.833)%
  --(3.746,1.522)--(4.293,1.393)--(4.839,1.237)--(5.386,1.211)--(5.933,1.160)--(6.480,0.978)%
  --(7.026,0.952)--(7.573,0.927)--(8.120,0.875)--(8.667,0.849)--(9.213,0.849)--(9.760,0.823)%
  --(10.307,0.823)--(10.854,0.771)--(11.400,0.771)--(11.947,0.745);
\gpsetpointsize{4.00}
\gppoint{gp mark 1}{(1.559,6.103)}
\gppoint{gp mark 1}{(2.106,3.437)}
\gppoint{gp mark 1}{(2.652,2.273)}
\gppoint{gp mark 1}{(3.199,1.833)}
\gppoint{gp mark 1}{(3.746,1.522)}
\gppoint{gp mark 1}{(4.293,1.393)}
\gppoint{gp mark 1}{(4.839,1.237)}
\gppoint{gp mark 1}{(5.386,1.211)}
\gppoint{gp mark 1}{(5.933,1.160)}
\gppoint{gp mark 1}{(6.480,0.978)}
\gppoint{gp mark 1}{(7.026,0.952)}
\gppoint{gp mark 1}{(7.573,0.927)}
\gppoint{gp mark 1}{(8.120,0.875)}
\gppoint{gp mark 1}{(8.667,0.849)}
\gppoint{gp mark 1}{(9.213,0.849)}
\gppoint{gp mark 1}{(9.760,0.823)}
\gppoint{gp mark 1}{(10.307,0.823)}
\gppoint{gp mark 1}{(10.854,0.771)}
\gppoint{gp mark 1}{(11.400,0.771)}
\gppoint{gp mark 1}{(11.947,0.745)}
\gppoint{gp mark 1}{(11.121,8.047)}
\gpcolor{color=gp lt color border}
\node[gp node right] at (10.479,7.539) {run 2};
\gpcolor{rgb color={0.580,0.000,0.827}}
\draw[gp path] (10.663,7.539)--(11.579,7.539);
\draw[gp path] (1.537,8.381)--(1.559,6.880)--(2.106,4.033)--(2.652,2.454)--(3.199,1.962)%
  --(3.746,1.625)--(4.293,1.496)--(4.839,1.367)--(5.386,1.237)--(5.933,1.160)--(6.480,1.082)%
  --(7.026,1.030)--(7.573,1.004)--(8.120,0.927)--(8.667,0.901)--(9.213,0.849)--(9.760,0.849)%
  --(10.307,0.849)--(10.854,0.797)--(11.400,0.745)--(11.947,0.745);
\gppoint{gp mark 2}{(1.559,6.880)}
\gppoint{gp mark 2}{(2.106,4.033)}
\gppoint{gp mark 2}{(2.652,2.454)}
\gppoint{gp mark 2}{(3.199,1.962)}
\gppoint{gp mark 2}{(3.746,1.625)}
\gppoint{gp mark 2}{(4.293,1.496)}
\gppoint{gp mark 2}{(4.839,1.367)}
\gppoint{gp mark 2}{(5.386,1.237)}
\gppoint{gp mark 2}{(5.933,1.160)}
\gppoint{gp mark 2}{(6.480,1.082)}
\gppoint{gp mark 2}{(7.026,1.030)}
\gppoint{gp mark 2}{(7.573,1.004)}
\gppoint{gp mark 2}{(8.120,0.927)}
\gppoint{gp mark 2}{(8.667,0.901)}
\gppoint{gp mark 2}{(9.213,0.849)}
\gppoint{gp mark 2}{(9.760,0.849)}
\gppoint{gp mark 2}{(10.307,0.849)}
\gppoint{gp mark 2}{(10.854,0.797)}
\gppoint{gp mark 2}{(11.400,0.745)}
\gppoint{gp mark 2}{(11.947,0.745)}
\gppoint{gp mark 2}{(11.121,7.539)}
\gpcolor{color=gp lt color border}
\node[gp node right] at (10.479,7.031) {run 3};
\gpcolor{rgb color={0.580,0.000,0.827}}
\draw[gp path] (10.663,7.031)--(11.579,7.031);
\draw[gp path] (1.714,8.381)--(2.106,4.136)--(2.652,2.557)--(3.199,2.091)--(3.746,1.781)%
  --(4.293,1.522)--(4.839,1.315)--(5.386,1.237)--(5.933,1.185)--(6.480,1.134)--(7.026,1.082)%
  --(7.573,1.030)--(8.120,1.004)--(8.667,0.978)--(9.213,0.952)--(9.760,0.901)--(10.307,0.901)%
  --(10.854,0.901)--(11.400,0.875)--(11.947,0.875);
\gppoint{gp mark 3}{(2.106,4.136)}
\gppoint{gp mark 3}{(2.652,2.557)}
\gppoint{gp mark 3}{(3.199,2.091)}
\gppoint{gp mark 3}{(3.746,1.781)}
\gppoint{gp mark 3}{(4.293,1.522)}
\gppoint{gp mark 3}{(4.839,1.315)}
\gppoint{gp mark 3}{(5.386,1.237)}
\gppoint{gp mark 3}{(5.933,1.185)}
\gppoint{gp mark 3}{(6.480,1.134)}
\gppoint{gp mark 3}{(7.026,1.082)}
\gppoint{gp mark 3}{(7.573,1.030)}
\gppoint{gp mark 3}{(8.120,1.004)}
\gppoint{gp mark 3}{(8.667,0.978)}
\gppoint{gp mark 3}{(9.213,0.952)}
\gppoint{gp mark 3}{(9.760,0.901)}
\gppoint{gp mark 3}{(10.307,0.901)}
\gppoint{gp mark 3}{(10.854,0.901)}
\gppoint{gp mark 3}{(11.400,0.875)}
\gppoint{gp mark 3}{(11.947,0.875)}
\gppoint{gp mark 3}{(11.121,7.031)}
\gpcolor{color=gp lt color border}
\node[gp node right] at (10.479,6.523) {run 4};
\gpcolor{rgb color={0.580,0.000,0.827}}
\draw[gp path] (10.663,6.523)--(11.579,6.523);
\draw[gp path] (1.536,8.381)--(1.559,6.802)--(2.106,3.567)--(2.652,2.247)--(3.199,1.936)%
  --(3.746,1.755)--(4.293,1.522)--(4.839,1.418)--(5.386,1.211)--(5.933,1.082)--(6.480,0.927)%
  --(7.026,0.875)--(7.573,0.797)--(8.120,0.797)--(8.667,0.771)--(9.213,0.745)--(9.760,0.694)%
  --(10.307,0.668)--(10.854,0.668)--(11.400,0.668)--(11.947,0.668);
\gppoint{gp mark 4}{(1.559,6.802)}
\gppoint{gp mark 4}{(2.106,3.567)}
\gppoint{gp mark 4}{(2.652,2.247)}
\gppoint{gp mark 4}{(3.199,1.936)}
\gppoint{gp mark 4}{(3.746,1.755)}
\gppoint{gp mark 4}{(4.293,1.522)}
\gppoint{gp mark 4}{(4.839,1.418)}
\gppoint{gp mark 4}{(5.386,1.211)}
\gppoint{gp mark 4}{(5.933,1.082)}
\gppoint{gp mark 4}{(6.480,0.927)}
\gppoint{gp mark 4}{(7.026,0.875)}
\gppoint{gp mark 4}{(7.573,0.797)}
\gppoint{gp mark 4}{(8.120,0.797)}
\gppoint{gp mark 4}{(8.667,0.771)}
\gppoint{gp mark 4}{(9.213,0.745)}
\gppoint{gp mark 4}{(9.760,0.694)}
\gppoint{gp mark 4}{(10.307,0.668)}
\gppoint{gp mark 4}{(10.854,0.668)}
\gppoint{gp mark 4}{(11.400,0.668)}
\gppoint{gp mark 4}{(11.947,0.668)}
\gppoint{gp mark 4}{(11.121,6.523)}
\gpcolor{color=gp lt color border}
\node[gp node right] at (10.479,6.015) {run 5};
\gpcolor{rgb color={0.580,0.000,0.827}}
\draw[gp path] (10.663,6.015)--(11.579,6.015);
\draw[gp path] (1.534,8.381)--(1.559,6.647)--(2.106,3.800)--(2.652,2.713)--(3.199,2.324)%
  --(3.746,2.040)--(4.293,1.677)--(4.839,1.444)--(5.386,1.263)--(5.933,1.082)--(6.480,1.082)%
  --(7.026,1.030)--(7.573,0.978)--(8.120,0.901)--(8.667,0.901)--(9.213,0.901)--(9.760,0.849)%
  --(10.307,0.823)--(10.854,0.823)--(11.400,0.771)--(11.947,0.771);
\gppoint{gp mark 5}{(1.559,6.647)}
\gppoint{gp mark 5}{(2.106,3.800)}
\gppoint{gp mark 5}{(2.652,2.713)}
\gppoint{gp mark 5}{(3.199,2.324)}
\gppoint{gp mark 5}{(3.746,2.040)}
\gppoint{gp mark 5}{(4.293,1.677)}
\gppoint{gp mark 5}{(4.839,1.444)}
\gppoint{gp mark 5}{(5.386,1.263)}
\gppoint{gp mark 5}{(5.933,1.082)}
\gppoint{gp mark 5}{(6.480,1.082)}
\gppoint{gp mark 5}{(7.026,1.030)}
\gppoint{gp mark 5}{(7.573,0.978)}
\gppoint{gp mark 5}{(8.120,0.901)}
\gppoint{gp mark 5}{(8.667,0.901)}
\gppoint{gp mark 5}{(9.213,0.901)}
\gppoint{gp mark 5}{(9.760,0.849)}
\gppoint{gp mark 5}{(10.307,0.823)}
\gppoint{gp mark 5}{(10.854,0.823)}
\gppoint{gp mark 5}{(11.400,0.771)}
\gppoint{gp mark 5}{(11.947,0.771)}
\gppoint{gp mark 5}{(11.121,6.015)}
\gpcolor{color=gp lt color border}
\draw[gp path] (1.012,8.381)--(1.012,0.616)--(11.947,0.616)--(11.947,8.381)--cycle;
\gpdefrectangularnode{gp plot 1}{\pgfpoint{1.012cm}{0.616cm}}{\pgfpoint{11.947cm}{8.381cm}}
\end{tikzpicture}
\end{minipage}
\vspace{-10pt}
\caption{The size of subgraphs corresponding to the unsatisfiable cores when using the algorithms {\sf TrimFormulaPlain} (left) and {\sf TrimFormulaInteract} (right).}
\label{fig:interact}
\end{figure}

We also experimented with the two algorithms presented in Section~\ref{sec:trim}. Figure~\ref{fig:interact} shows the size of the subgraph (extracted from the core)
for the first $20$ iterations with formula $F^+_{4}$ as input using {\sf TrimFormulaPlain} (left) or {\sf TrimFormulaInteract} (right). Each experiment was run five times.
The figure shows that {\sf TrimFormulaInteract} produces significantly smaller subgraphs. The {\sf TrimFormulaPlain} algorithm, as shown in Fig.~\ref{fig:trim}, actually performs
significantly worse than the performance presented in Fig.~\ref{fig:interact}. This poor performance is caused by the removal of edge clauses and symmetry-breaking predicates
from the core. We improved the {\sf TrimFormulaPlain} algorithm by adding back the removed edge clauses and symmetry-breaking predicates in each iteration.

We studied the resulting graphs and observed that they were close to symmetric: Taking the union of the graph with rotated copies ($120$ degrees rotation in the origin) added only 
a few dozen vertices. We decided to check whether this observation could be used to further shrink the large part by taking this union as initial graph (instead of $G_{2167}$)
and rerun the procedure. This turned out to be effective and allowed removing some additional vertices. We ran the entire experiment many times on a cluster 
Several runs resulted in a graphs with ``only'' $393$ vertices. These graphs turned out to be the same (modulo rotation and reflection). 
We call this graph $L_{393}$. One can make $L_{393}$ symmetric, i.e., it maps onto itself when rotating it by 120 degrees along the central vertex, by adding a single vertex.

\subsection{Finalizing the Graph}

The graph $L_{393}$, produced in the previous subsection, needs to be extended with a ``small part'' to establish a unit-distance graph 
with chromatic number 5. Initially we tried to use the small part of $G_{553}$. However, the resulting graph is 
$4$-colorable, because $L_{393}$ has fewer connections with that small part compared to the large part of $G_{553}$. We fixed this as
follows: The small part of $G_{553}$ got expanded by merging it with copies that are $60$ degrees rotated in the origin. This resulted in a graph 
with $181$ vertices (while the small part of $G_{553}$ has $134$ vertices), which we call $S_{181}$. The union of $L_{393}$ and $S_{181}$ has
chromatic number $5$. We applied the same techniques as described in the previous subsection to further reduce the size of this 
graph. This resulted in a graph being the union of $L_{393}$ and a new small part with $137$ vertices. 

\begin{figure}[h!]
\centering
\includegraphics[width=.98\textwidth]{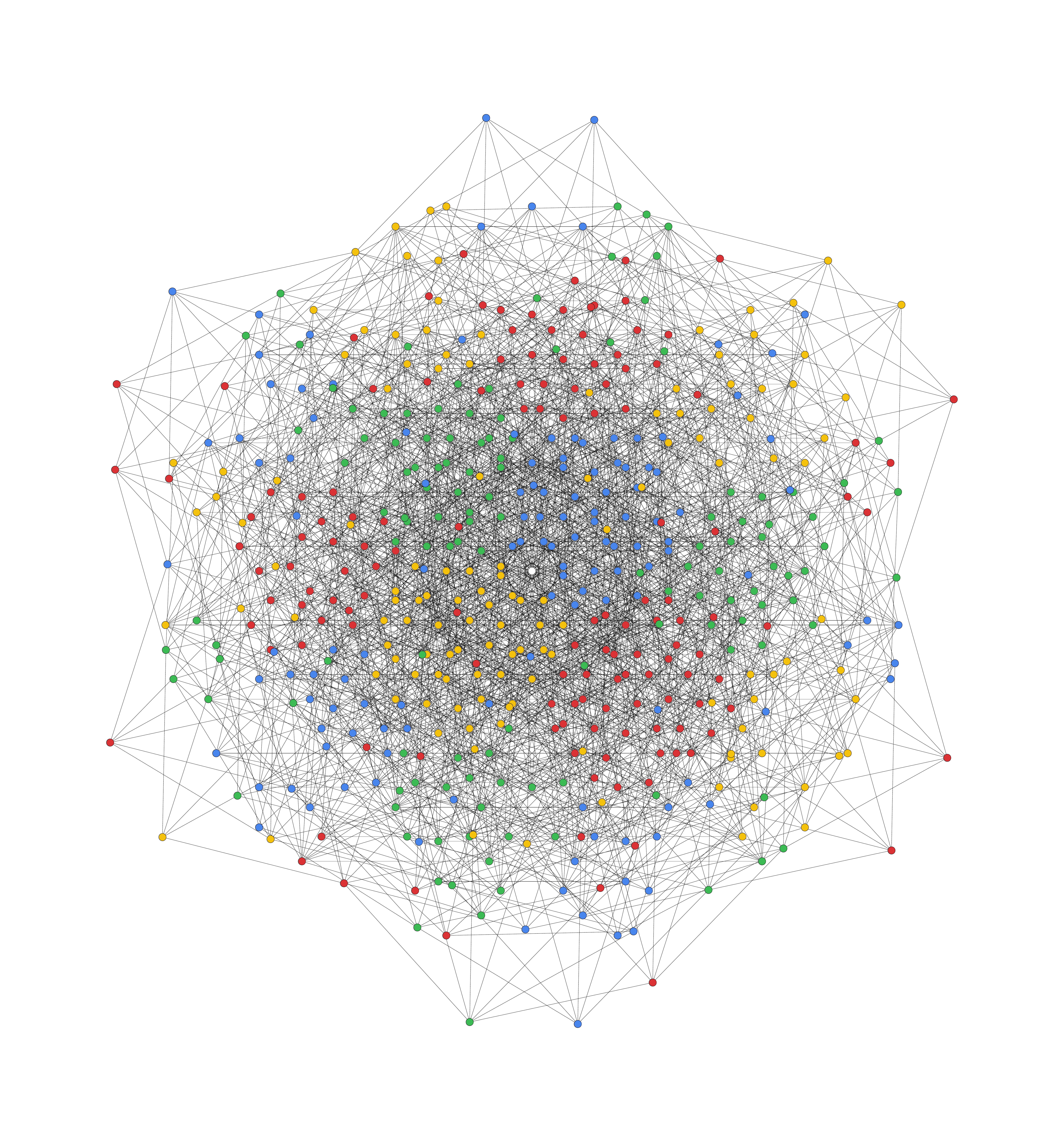}
\caption{A 529-vertex unit-distance graph with chromatic number 5. In the shown coloring, only the origin has the fifth color (white). \vspace{-10pt}}
\label{fig:W529}
\end{figure}

Figure~\ref{fig:W529} shows the final graph $G_{529}$ consisting of $529$ vertices and $2\,670$ edges.  This graph almost
maps onto itself when rotating it with $120$ degrees in the origin. The figure shows a coloring in which only the origin has the fifth 
color (white). Such a coloring exists for each vertex as the graph is vertex critical. The shown coloring is a randomly selected
one. Observe the clustering of vertices with the same color. This pattern looks similar to the one shown in Fig.~\ref{fig:Htwice}.

Graph $G_{529}$ is available at \url{https://github.com/marijnheule/CNP-SAT} as a list of points in the plane and a list of unit-distance edges.
The repository also contains a CNF formula encoding whether $G_{529}$ is 4-colorable and a proof of unsatisfiability that can be validated in a few seconds. 

\section{Conclusions}

We presented a new algorithm to trim a formula by first optimizing a proof of unsatisfiability. The algorithm optimizes the proof 
using both the shrinking formula and the original formula. This allows reintroducing clauses in the shrinking formula, which 
could further improve the trimming.

We constructed a unit-distance graph with points in $\mathbb{Q}[\sqrt{3}, \sqrt{11}] \times \mathbb{Q}[\sqrt{3}, \sqrt{11}]$. The
4-colorings of this graph, $G_{2167}$, have some interesting properties such as 1) many (and in some 4-colorings all) vertices with
the same horizontal coordinate have the same color; 2) vertices that are 1/3 apart having a different color; and 3) vertices with 
the same vertical coordinate that are 2/3 apart have the same color. All these properties are for a rotation of $G_{2167}$ by
0, 120, or 240 degrees.

By combining the new algorithm and the new graph, we were able to reduce the smallest known unit-distance graph with
chromatic number 5 to a graph with 529 vertices and 2670 edges (down from 553 vertices and 2720 edges).  This graph is also much more symmetric.
It is generally easier to understand why a symmetric object has a certain property
compared to an asymmetric object. It may thus provide some insight how to obtain a unit-distance graph with chromatic
number 6 (if they exist).  Using the techniques in the paper we constructed several graphs with up to $100\,000$ vertices,
but all were 5-colorable. 

As future work, we plan to study the effectiveness of the new algorithm on other applications that require minimal unsatisfiable cores.

\section*{Acknowledgements}

The author is supported by the National Science Foundation (NSF) under grant CCF-1813993.
The author acknowledges the Texas Advanced Computing Center (TACC) at The University of Texas at Austin for providing HPC
resources that have contributed to the research results reported within this paper. 


\bibliography{CNP}
\bibliographystyle{splncs03}

\end{document}